\documentclass[a4paper,11pt]{article}
\pdfoutput=1 

\usepackage{jcappub} 

\usepackage[T1]{fontenc} 

\usepackage{jcappub} 
\usepackage{natbib}  

\title{\boldmath A title with some math: $x=1$}

\usepackage[utf8]{inputenc}
\DeclareUnicodeCharacter{03B3}{\ensuremath{\gamma}}
\DeclareUnicodeCharacter{2212}{-}
\DeclareUnicodeCharacter{0306}{\breve{}}
\usepackage{amsmath}
\usepackage{float}

\usepackage{lscape}
\usepackage{subcaption}

\title{Determining energy limits for proton acceleration through high-energy $\gamma$-ray observations with CTAO}

\author[a,1]{P.~Sharma,\note{Corresponding author.}}
\emailAdd{pooja.sharma@thapar.edu}

\author[b]{C.~Dubos,}

\author[c]{S. R. Patel}

\author[b]{and~T.~Suomij\"arvi}
\emailAdd{tiina.suomijarvi@ijclab.in2p3.fr}

\affiliation[a]{Thapar Institute of Engineering \& Technology, Patiala, India}

\affiliation[b]{Universit\'{e} Paris-Saclay, CNRS/IN2P3, IJCLab,  \\ 15 rue Georges Clemenceau, 91405 Orsay, France}

\affiliation[c]{Helmholtz-Zentrum Berlin für materialien und energie,\\Hahn-Meitner-Platz 1, 14109 Berlin, Germany}

\abstract{This paper reports on the capabilities of the Cherenkov Telescope Array Observatory (CTAO) in defining the energy cut-off of the cosmic-ray proton distribution by observing high-energy $\gamma$-rays.
We focus on four sources: Cassiopeia A, HAWC J2227+610, HESS J1731-347 and RX J1713.7-3946, which have been previously identified 
to have an important hadronic contribution extending to the highest energies in the $\gamma$-ray spectra.
Simulations were performed using \textsc{Gammapy} for each source to obtain simulated spectral energy distributions for CTAO. In the case of HAWC J2227+610, a detailed study was performed to define the maximum cut-off energy for proton distribution that would be detectable by measuring $\gamma$-rays with CTAO. To distinguish between fluxes with different proton cut-off energies, we used the Test Statistic (TS) method. This study showed that, in the case of HAWC J2227+610, CTAO would be able to detect proton cutoff energies up to about $\sim$ 600\,TeV. }


\begin{document}
\maketitle\flushbottom

\section{Introduction}
The Cosmic Rays (CRs) below the knee ($\sim$ PeV energies) of the CR energy spectrum are expected to have a Galactic origin. Objects such as Supernova Remnants (SNRs), Pulsar Wind Nebulae (PWNe), or stellar clusters are widely regarded as likely sites for CR acceleration, supported by strong theoretical and observational evidence \cite{2013_Blasi}. However, to identify PeVatrons, sources capable of accelerating hadrons up to PeV energies, we require a precise understanding of the maximum energies to which these sources can accelerate particles, which in turn necessitates robust observational evidence. At such energies, it is difficult to point back to the source of CRs directly, as these CRs are charged particles and are deflected by the magnetic fields in the Galaxy, causing them to deviate from their initial path. One of the primary methods to confirm hadronic acceleration is the detection of very high-energy (VHE) $\gamma$-rays. As neutral messengers, $\gamma$-rays can travel undeflected through magnetic fields, providing a direct means to trace back to their sources. The interaction of the accelerated CRs with the surrounding medium can lead to the emission of $\gamma$-rays through the inelastic production of neutral pions $\pi^0$ in proton-proton collisions;
\begin{equation}
\left\{
\begin{array}{l}
  p + p \to \pi^0 + p + p \\
  \pi^0 \to 2\gamma
\end{array}
\right.
\end{equation}
In addition, some other channels can contribute, such as interactions involving heavier nuclei and the production of intermediate baryonic resonances (e.g. $\Delta$ or $N^*$), which subsequently decay and enhance the overall and $\gamma$-ray yield.
The $\gamma$-ray emission spectra resulting from the pion-decay process follow the parent proton spectrum, exhibiting a cut-off at energies typically one order of magnitude lower than the maximum proton energy. Therefore, the presence of PeV protons is inferred indirectly through the detection of $\gamma$-rays extending to $\sim$ 100 TeV \cite{kelner2006energy}.
\noindent
We have witnessed significant discoveries by the current generation of $\gamma$-ray instruments such as H.E.S.S., MAGIC, VERITAS, HAWC, and notably LHAASO. Several sources have been identified as PeVatron candidates, owing to the detection of $\gamma$-rays beyond 100 TeV \cite{hess2016acceleration}, \citep{2021Natur}. Recent observation of $\gamma$-rays from the SNR G$106.3+2.7$ (commonly known as HAWC J2227+610) at very high-energies above 100\,TeV has gained popularity as a PeVatron candidate \cite{Albert_2020}. Despite these breakthroughs, analyzing the cut-off in the parent proton spectrum has been challenging due to limitations in sensitivity and angular resolution of the current instruments. This limitation critically affects the capacity to discriminate between purely leptonic and lepto-hadronic emission scenarios at very high energies. Constraining the proton energy cut-off is pivotal for establishing the source as a PeVatron and for probing the nature of the underlying cosmic-ray acceleration processes \cite{malkov2001nonlinear}, \cite{cristofari2021hunt}.\\

\noindent
Great leaps in $\gamma$-ray astronomy are expected by the advent of Cherenkov Telescope Array Observatory (CTAO), the next-generation $\gamma$-ray observatory \cite{acharya2017science}. CTAO will be the largest ground-based Imaging Atmospheric Cherenkov Telescope (IACT) observatory. In its Alpha configuration, which includes 14 Medium-Sized Telescopes (MSTs) and 37 Small-Sized Telescopes (SSTs) in the southern site and 4 Large-Sized Telescopes (LSTs) and 9 MSTs in the northern site, CTAO will be able to detect $\gamma$-rays with a sensitivity 10x better than its predecessors over an energy range of 30\,GeV to 200\,TeV. With a field of view of up to 10$^{^{\circ}}$, its dual-site configuration in both hemispheres will give access to the entire sky. CTAO will allow us to perform an unprecedented study of particle acceleration in Galactic sources \cite{lopez2025ctao}. In a recent paper \cite{ACERO_pev_CTA}, authors use CTAO Galactic Plane Survey (GPS) simulated data to show that CTAO can confirm the PeVatron nature of SNRs given sufficient exposure \cite{abe2024prospects}. The recent multi-wavelength (MWL) study of the ultra-high-energy $\gamma$-ray source LHAASO J2108+5157 by XMM-Newton, Fermi-LAT, and LST-1 \cite{abe2023multiwavelength} further highlights the importance of using data spanning a wide wavelength range to disentangle the leptonic vs lepto-hadronic nature of source emission. While still in its initial phase, LST-1 has demonstrated promising performance and sensitivity, proving its potential to provide robust observational constraints and help reduce parameter space degeneracy in future CTAO observations. While CTAO’s angular resolution and sensitivity will indeed be crucial for disentangling leptonic and hadronic contributions in general, the present work focuses specifically on the spectral domain. In particular, we investigate how well CTAO observations can constrain the proton energy cut-off in the $\gamma$-ray spectra, which provides a direct probe of the maximum energies reached in Galactic accelerators. \\ 

\noindent
This study expands upon our earlier published work on the MWL analysis of Galactic SNRs \cite{SNR_Sharma}, hereafter referred to as MWL23. In the paper MWL23, we compiled observed Spectral Energy Distribution (SED) from various well-cited publicly available MWL data of 9 SNRs, as listed in Table 7 of Appendix A of MWL23. We also studied various parent particle distributions, including a single power law, a broken power law, and an exponentially cut-off power law (ECPL). Among these, the ECPL model provided the best fit to the observed MWL data, which motivated its use in our analysis. Finally, the observed MWL SED for each source was fitted using several theoretical high-energy emission models implemented in \textsc{Naima} package \cite{Naima_zabalza}. Naima employed the MCMC method \cite{Foreman_Mackey_2013} to fit the observations with the theoretical models. Naima's fitting method required an initial set of parameters, for which a broad range of initial parameters was derived using existing literature. After multiple initializations and convergence checks, the best-fit parameters were obtained by minimizing the likelihood, thereby representing the statistically preferred values under the assumptions made. Through likelihood comparisons and evaluation via the Bayesian Information Criterion (BIC), our research showed that the lepto-hadronic scenario was consistently favored across all SNRs. Among these, four sources - Cassiopeia A, HAWC J2227+610, HESS J1731-347, and RX J1713.7-3946 stood out with strong indications of hadronic contributions up to the highest energies, marking them as promising PeVatron candidates.

The summary of the current paper is as follows: Section 2 presents a detailed description of the four SNRs selected for this study. In Section 3, we outline the modeling of the expected $\gamma$-ray flux as it would be observed by CTAO, assuming pion-decay emission resulting from the parent proton distribution. In section 4, we then assess CTAO’s capability to observe these sources with improved flux precision over a wide energy range by comparing simulated data with existing MWL observations. Finally, Section 5 focuses on HAWC J2227+610, evaluating CTAO’s sensitivity to different proton energy cut-offs. 


\section{Selected sources}
For the selected sources, basic information obtained from the TeVCat database \cite{articletevcat} and the literature is summarized in Table \ref{tab:TypeSources}. The selected sources are shell-type and interacting SNRs, their distances vary from 0.8\,kpc to 3.4\,kpc, and ages range from 340 years to 6000 years. The reported magnetic fields lie between 14 to 400\,$\mu G$. The sources RX J1713.7-3946 and HESS J1731-347 are visible by CTAO South. HAWC J2227+610 and Cassiopeia A are visible by CTAO North.

\begin{table}[H]
\caption{\label{tab:TypeSources}List of the chosen SNRs along with basic information found from TeVcat and literature. Here, Int. refers to SNRs interacting with the surrounding medium and Shell to shell-type SNRs.}
\scriptsize
\hspace{-1cm}
\begin{tabular}{|l|c|c|c|c|c|c|c|c|c|}
\hline
Source & RA & DEC & Gal Long & Gal Lat & Dist. & Spec. Index & Type & Age & $B$ ($\mu$G) \\
       & (hh mm ss) & (dd mm ss) & (deg) & (deg) & (kpc) &  &  & (yrs) &  \\
\hline
Cassiopeia A & 23 23 13.8 & +58 48 26 & 111.71 & -2.13 & 3.4 & 2.3 & Shell & 340~\cite{CasA_1999} & 400~\cite{helder2008characterizing} \\
HAWC J2227+610  & 22 27 59 & +60 52 37 & 106.35 & 2.71 & 0.8 & 2.29 & Int. & 4700--5700~\cite{Albert_2020} & 100~\cite{bao2021hard} \\
HESS J1731-347  & 17 32 03 & -34 45 18 & 353.54 & -0.67 & 3.2 & 2.32 & Shell & 2000--6000~\cite{cui2019snr} & 50~\cite{abramowski2011new} \\
RX J1713.7-3946 & 17 13 33.6 & -39 45 36 & 347.34 & -0.47 & 1 & 2.2 & Shell & 1580--2100~\cite{tsuji2016expansion} & 14~\cite{abdalla2018hess} \\
\hline
\end{tabular}
\end{table}

\noindent
These sources have been the focus of extensive prior investigations due to their potential role as CR accelerators. A concise overview of these studies is provided below.

\subsection{Cassiopeia A}

Cassiopeia A is a bright, young shell-type SNR that has been proposed to be a site of CR acceleration \cite{cassA_gamma_evidence}. Hadronic emission from a component of the SNR has been found through the MWL studies \cite{Abeysekara_2020_cassA}. Although there is strong evidence for hadronic acceleration up to the highest energies, its observed spectral cut-off around a few TeV is well below the PeV range, making it unlikely to be a PeVatron capable of accelerating cosmic rays to the knee of the Galactic spectrum ($\sim$1 PeV). 

\subsection{HAWC J2227+610}
HAWC J2227+610 is currently one of the most promising PeVatron SNR candidates. It has shown $\gamma$-ray emission up to more than 100 TeV, with no hint of cut-off in its high-energy spectrum \cite{Albert_2020}. Thus, this makes it an interesting target for CTAO. It is expected to be out of the field of view of CTAO South, which has been optimized for the study of high-energy Galactic sources. However, CTAO North can observe it with high sensitivity by ensuring a longer observing time. G. Verna et al. \cite{Verna:2021} explored multiple CTAO array layouts (small and large) and integration times (ranging from 50 to 200 hours), showing that CTAO could detect the source with a significance exceeding 5$\sigma$.  For a hadronic emission model, CTA-North can constrain the cut-off energy of HAWC J2227+610 to within $\sim$14\% (small array) to ~10\% (large array) after 200 hours of observation, with even tighter constraints ($\sim$1\%) on the amplitude and spectral index. Despite this, the study concluded that even with 200 hours of exposure, CTAO alone would not be able to discriminate between hadronic and leptonic emission scenarios due to spectral degeneracies at TeV energies. The small and large arrays explored in earlier CTAO studies represent design extremes in sensitivity and telescope density, whereas the Alpha configuration corresponds to the finalized baseline CTAO layout, offering a balanced and realistic performance across the full energy range.

\subsection{HESS J1731-347}

HESS J1731-347 is a young SNR whose MWL counterparts have been studied greatly since its discovery \cite{2011A&A_hess_j1731}. Its proximity to a nearby molecular cloud makes this SNR an interesting object of study \cite{2019_hessj1731_MCs}. However, MWL SED fitting of HESS J1731−347 indicates a relatively low proton cut-off energy of $\sim$20 TeV \cite{SNR_Sharma}, which is well below the PeV scale, making it unlikely to be a PeVatron candidate.

\subsection{RX J1713.7-3946}
Evidence for TeV $\gamma$-ray emission from the SNR RX J1713.7-3946 was established by \cite{rxj17132000evidence}. After that, this SNR has become one of the best-studied young SNRs. A paper by the H.E.S.S. Collaboration \cite{rxj1713_hess_2018} showed that the broadband spectra ranging from radio to $\gamma$-rays could not conclusively confirm whether the emission is of leptonic or hadronic nature or a mix of both. Furthermore, simulations for CTAO observations of RX J1713.7−3946 were performed, and they revealed that with the exposure time of 50 hours, CTAO would be able to identify the dominant $\gamma$-ray emission component from the morphology of the SNR \cite{Acero_2017_rxj1713_cta}.

\section{Simulations for CTAO}

For each of these selected SNRs, we have modeled the expected $\gamma$-ray SED from hadronic interactions using $\pi^0$-decay emission. The simulation requires assuming several parameters, such as the density of protons in the surrounding medium $N_H$ and the underlying particle distribution. These parameters have been adopted from our previous work, MWL23. Simulations were performed using \textsc{Gammapy} \cite{2019A&A...625A..10N}. Gammapy is a Python package that allows simulating and analyzing $\gamma$-ray data. To model the expected hadronic emission, we used the Naima library to compute the $\gamma$-ray flux based on radiative models.



\subsection{Proton distribution}

Since our study focuses on hadronic emission, we have considered protons for the parent particle distribution. We assume the energy distribution of protons, $f_p(E)$, to follow an exponential cut-off power law (ECPL). The distribution for protons is given below:
\begin{align}  f_p(E) & = A_p \times \left(\frac{E}{E_0}\right)^{-\alpha_p} \times \exp ({-E/E_{p,cut})^ \beta} \label{eq:1} \end{align}
with:
\begin{itemize}
    \item $A_p$: Amplitude of the proton distribution
    \item $\alpha_{p}$: Proton spectral index
    \item $E_{p,cut}$: Proton energy cut-off
    \item $\beta$: Cut-off exponent
\end{itemize}

\noindent
The resulting photon flux from the proton distribution is directly proportional to the gas
density ($N_H$) and the inelastic cross-section between CR protons and stationary protons from the surrounding material $\sigma$. However, it should be noted that the connection between the underlying parent particle distribution and the observed $\gamma$-ray emission is non-trivial. Naima implements a detailed semi-analytical parameterization of the $\pi^0$ emission spectrum defined by Kafexhiu et al. \cite{2014PhRvD..90l3014K} based on experimental data of proton-proton collisions. It is important to note that the parametrization introduces a non-negligible level of uncertainty of $\sim$20\%, across a broad energy range from the kinematic threshold up to PeV energies. This uncertainty arises from combining different theoretical models across energy regimes, limitations in the available experimental cross-section data (especially at high energies), and differences in the Monte Carlo codes used to simulate hadronic interactions. Accounting for this uncertainty is beyond the scope of this paper.

\subsection{Radiative models}

In the previous work, MWL23, the following radiative models were considered: synchrotron, bremsstrahlung, Inverse Compton (IC), and pion-decay. A lepto-hadronic scenario (synchrotron, bremsstrahlung, IC, and pion-decay) and a pure leptonic scenario (synchrotron, bremsstrahlung, IC) were taken into consideration to model the expected flux. The results of the SED fitting showed that the lepto-hadronic scenario was preferred compared to the leptonic scenario. The integrated flux across energy bins for the chosen SNRs, ranging from $ 10^{10}$ to $10^{12}$ eV, showed that the hadronic component (specifically pion-decay) dominates the emission, contributing over 85\% and exceeding the flux contribution due to leptonic processes by more than an order of magnitude. Therefore, for CTAO simulations, only the hadronic model with pion-decay was considered. In addition, using the pion-decay model significantly decreases the number of parameters compared to combining other radiative models.\\

\noindent
The reference energy was set to $E_0 = 1$\,TeV and $\beta = 1$. 

\begin{table}[H]
    \centering
    \begin{tabular}{ |p{2.2cm}|p{2.2cm}|p{2.2cm}|p{2.2cm}|p{2.2cm}| }
      \hline
Parameters       & Cassiopeia A  & HAWC J2227+610 & HESS J1731-347 &  RX J1713.7-3946 \\
      \hline
      log10(A) & $47.03\substack{+0.0002 \\ -0.0004}$ & $47.02\substack{+0.001 \\ -0.001}$ & $47.28\substack{+0.007 \\ -0.009}$ &   $49.90\substack{+0.02 \\ -0.02}$ \\
      \hline
      $\alpha_{p}$  & $2.10\substack{+0.01 \\ -0.02}$ & $1.76\substack{+0.02 \\ -0.03}$ & $1.64\substack{+0.04 \\ -0.04}$ &  $1.75\substack{+0.02 \\ -0.01}$\\
      \hline
      $E_{p,cut}$(TeV) & $23.40\substack{+0.01 \\ -0.01}$ & $446.68\substack{+0.07 \\ -0.07}$ & $20.40\substack{+0.04 \\ -0.06}$ &  $74.04\substack{+0.04 \\ -0.05}$ \\
      \hline
      $N_H$ ($\text{cm}^{-3}$) & $163.21\substack{+1.43 \\ -2.37}$ &  $1.70\substack{+0.09 \\ -0.10}$ & $45.16\substack{+2.04 \\ -2.16}$ &  $10.29\substack{+0.15 \\ -0.16}$ \\
      \hline
    \end{tabular}
    \caption{\label{tab:param_out_naima}Hadronic model parameters from the previous MWL study \cite{SNR_Sharma}.}
\end{table}

\noindent
Table \ref{tab:param_out_naima} presents the values of the pion-decay model parameters derived from the paper MWL23, which serve as the basis for the simulations conducted in this study.

We note that some of the parameters lie at the edge of physically acceptable values. Specifically, the value $N_H$ for Cassiopeia A is quite high, however, Naima also provides the value of the average cosmic-ray density, $W_P (E_p > 1 TeV)$, which is 3.86 $ \times 10^{47}$ ergs. If the canonical kinetic energy of a supernova explosion is $\sim$ $10^{51}$ ergs, and we assume 10\% was used to accelerate CRs, then the derived fit parameter falls well within the allowed CR budget. Since $\gamma$-ray luminosity, $L_{\gamma}$ $\propto$ $N_H \cdot \,W_p \cdot \, V$, where V is the volume, a high value of $N_H$ implies a large amount of target material for CR interactions. However, it is difficult to isolate, without external constraints, the degeneracy with cosmic-ray density. The presence of a nearby molecular cloud further supports the $N_H$ value \cite{kilpatrick2014interaction}. 
Furthermore, spectral indices below 2, such as those observed for HAWC J2227+610 and HESS J1731-347, are harder than what is typically predicted by standard diffusive shock acceleration (DSA) theory. However, such values are supported in the literature \cite{2021}, \cite{condon2017detection}. The measured index, $\alpha$, lies between the canonical DSA prediction of $\alpha = 2$ and the theoretical lower limit of highly efficient proton acceleration, $\alpha = 1.5$ \cite{malkov2001nonlinear}.


\subsection{Simulations with \textsc{Gammapy}}
\label{3.3}

The flux simulations were performed using GAMMAPY, which allows us to generate $\gamma$-ray datasets by combining source emission models with Instrument Response Functions (IRFs) of CTAO. For this study, we have used \textit{prod5 version v0.1} which provides IRFs \cite{ctao_IRF} for both sites (Southern or Northern), for 3 different zenith angles of the source (20\textdegree, 40\textdegree \,or 60\textdegree) and for 3 different observation times (0.5\,h, 5\,h or 50\,h). We have used a zenith angle of 40\textdegree \,for all sources. In this work, we focus exclusively on spectral modeling, where the source emission is described by a physically motivated proton distribution that generates $\gamma$-rays through pion-decay. The simulated datasets were analyzed to extract fluxes and detection significances using likelihood-based estimators within Gammapy. The uncertainties reported in this work reflect statistical errors only, derived from the likelihood analysis of simulated CTA observations, and do not include instrumental or modeling systematic uncertainties. Consequently, these uncertainties are smaller than those typically reported in the literature, where systematic effects are often included. Accounting for such systematics will be a key challenge for parameter estimation once CTAO is built. The uncertainties reported for CTAO simulations represent the dispersion across multiple realizations and should not be directly compared to single-realization likelihood errors from MWL studies using Naima.

\noindent
The observation parameters for simulations are tabulated in Table \ref{tab:sim_params}.

\begin{table}[H]\small
    \centering
    \begin{tabular}{ |p{3.5cm}|p{2.4cm}|p{2.4cm}|p{2.4cm}|p{2.4cm}| }
      \hline
      \textbf{Parameter} & \textbf{Cassiopeia A}  & \textbf{HAWC J2227+610} & \textbf{HESS J1731-347} & \textbf{RX J1713.7-3946} \\
      \hline
      IRF & CTAO North Prod5 & CTAO North Prod5 & CTAO South Prod5 & CTAO South Prod5 \\
      \hline
      Zenith Angle ($^\circ$) & 40 & 40 & 40 & 40 \\
      \hline
      Acceptance ratio $\alpha$ & 0.2 & 0.2 & 0.2 & 0.2 \\
      \hline
      Galactic Longitude ($^\circ$) & 111.71 & 106.35 & 353.54  &  347.34 \\
      \hline
      Galactic Latitude ($^\circ$) & -2.13  & 2.71  & -0.67  & -0.47 \\
      \hline
      Distance (kpc) & 3.4 & 0.8 & 3.2 & 1.0 \\
      \hline
      Source Type & Point & Extended & Extended & Extended \\
      \hline
    \end{tabular}
    \caption{\label{tab:sim_params}Basic simulation parameters used for simulating CTAO fluxes.}
\end{table}

\noindent

We have simulated the CTAO spectrum using synthetic On-Off spectral analysis, where the background has been simulated by generating Poisson-distributed counts in both the signal (On) and background (Off) regions using the model-predicted background.
For the simulation, an acceptance ratio of $\alpha = $ 0.2 has been assumed, which is the ratio of the area scaling of the on-region to the area scaling of the off-region. The chosen value of $\alpha$ corresponds to an On-region of 1 and an Off-region area of 5. This provides us with consistent background scaling across all realizations while avoiding spatial systematics. The simulation parameters are tabulated in the table \ref{tab:sim_params}.

While performing simulations, it is important to study the statistical fluctuations in the simulated dataset. Such fluctuations, also called the sample variance, need to be accounted for when interpreting the reconstructed parameters and significance when comparing different models. To take sample variance into account, for our study, we simulated N independent datasets, and each was fitted using the appropriate model for a specific study. For the fit, we used the optimizer \textsc{Minuit} \cite{James:1975dr}, which can be accessed through \textsc{Gammapy}'s \textit{Fit class}. Depending on the study, the cut-off energy was either left free (precision study) or fixed (model discrimination). From the ensemble of fits, we derived mean parameter values, standard deviations, and Test Statistic (TS) distributions, which allowed us to quantify the precision ($\sigma$/mean) of the reconstructed parameters and significance. These simulations for RX J1713.7-3946, HAWC J2227+610, and HESS J1731-347 were performed as extended sources. The 1D analysis approach was chosen because the studied SNRs are isolated sources. Exclusion masks were used to prevent contamination of the background estimate from $\gamma$-ray bright regions, as identified using \href{http://gamma-sky.net/}{gamma-sky.net}. However, these exclusion masks are not particularly necessary for the studied sources, as the other $\gamma$-ray sources in the region do not overlap with them.\\





\section{Study of simulated fluxes}

As a first step, we simulated CTAO SED fluxes for all four SNRs using the methodology and parameters outlined above to assess the additional constraints that CTAO observations can provide. 

\subsection{Precision study of reconstructed parameters}

\begin{figure}[H]
    \hspace{-1.5cm} 
    \begin{minipage}{0.48\textwidth}
        \includegraphics[scale=0.40]{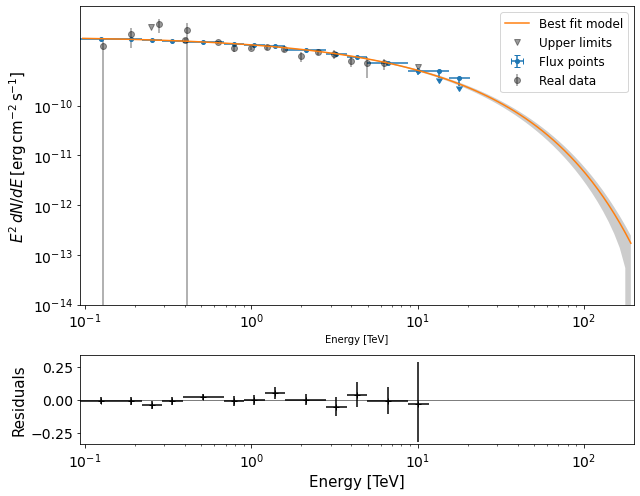}
    \end{minipage}
    \hspace{0.1\textwidth}
    \begin{minipage}{0.48\textwidth}
        \includegraphics[scale=0.40]{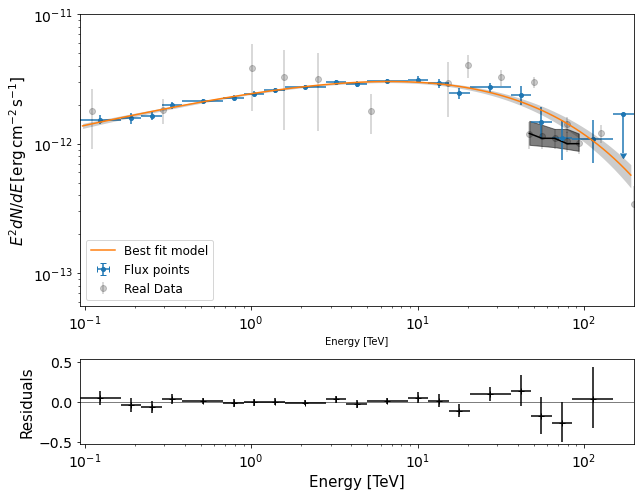}
    \end{minipage}
    \caption{CTAO simulated flux for Cassiopeia A (left) and HAWC J2227+610 (right). The simulated fluxes (blue) are generated by the theoretical pion-decay model (orange curve). For comparison, the grey points indicate the real data obtained from \textit{Fermi}-LAT \cite{Yuan_2013}, MAGIC \cite{Albert_2007} and VERITAS \cite{Acciari_2010} for Cassiopeia A and from \textit{Fermi}-LAT \cite{2019ApJ...885..162X}, VERITAS \cite{Acciari_2009}, LHAASO \cite{2021Natur.594...33C} and HAWC \cite{Albert_2020} for HAWC J2227+610. The simulated flux shown corresponds to a single realization of the model.}
    \label{SNR_3_4}
\end{figure}
\begin{figure}[H]
    \hspace{-1.5cm} 
    \begin{minipage}{0.48\textwidth}
        \includegraphics[scale=0.40]{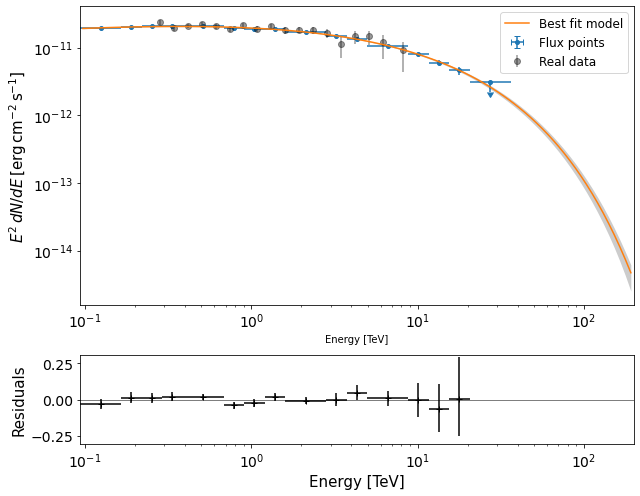}
    \end{minipage}
    \hspace{0.1\textwidth}
    \begin{minipage}{0.48\textwidth}
        \includegraphics[scale=0.40]{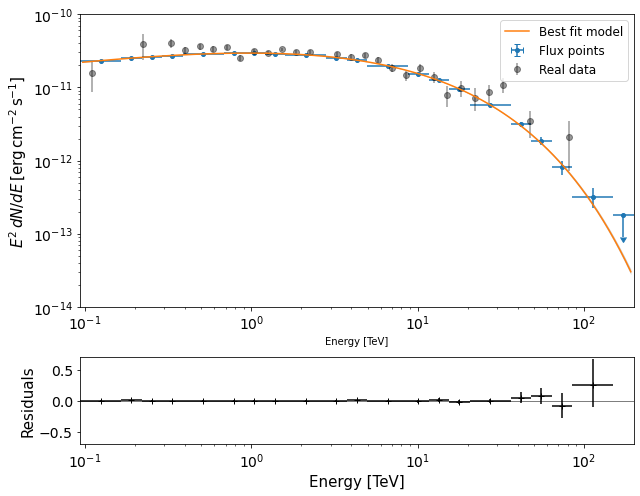}
    \end{minipage}
    \caption{ CTAO simulated flux for HESS J1731-347 (left) and RX J1713.7-3946 (right). The simulated fluxes (blue) are generated by the theoretical pion-decay model (orange curve). For comparison, the grey points indicate the real data obtained from H.E.S.S \cite{2011} for HESS J1731-347 and \textit{Fermi} \cite{Abdo_2011} and H.E.S.S \cite{2007A&A...464..235A} for RX J1713.7-3946. The simulated flux shown corresponds to a single realization of the model.}
    \label{SNR_1_2}
\end{figure}

We simulated 50 datasets with an exposure of 100 hrs, and $\gamma$-ray fluxes were simulated based pion-decay emission for all 4 sources. These datasets were then fitted using the corresponding input spectral model, allowing the proton cut-off energy, spectral index, and amplitude to vary freely in order to evaluate how accurately these parameters could be reconstructed relative to their injected values.

Figure \ref{SNR_3_4} and \ref{SNR_1_2} show the simulated SED for one of the realisations, in the energy range of 30\,GeV - 199\,TeV. The simulated CTAO data points are compared to the theoretical pion-decay model and to real data points from different experiments (see figure captions). By comparing the simulated CTAO data and the real data, one can observe that the CTAO flux extends to larger energies than that of the current IACTs. We find that the statistical uncertainties on the flux measurements using simulated CTAO data are reduced by approximately an order of magnitude compared to current real observations, across all sources and in a given energy range. The difference in flux errors between simulated and real data points is about a factor of 10. This improvement would allow for better constraints on the fit parameters and a more accurate determination of the proton cut-off energy.

\begin{table}[H]\small
    \centering
    \begin{tabular}{ |p{3.5cm}|p{3.2cm}|p{3.2cm}|p{3.2cm}| }
      \hline
      \textbf{Source} & \textbf{$E_{\rm p,cut}$ [TeV]} & \textbf{ Spectral Index $\alpha_p$} & \textbf{Amplitude} \\
      \hline
      Cassiopeia A & $23.5 \pm 2.4$ (10.3\%) & $2.10 \pm 0.026$ (1.2\%) & $(1.07 \times 10^{47}) \pm (1.5 \times 10^{45})$ (1.4\%) \\
      \hline
      HAWC J2227+610 & $452 \pm 72.8$ (16.3\%) & $1.76 \pm 0.0268$ (1.5\%) & $(1.05 \times 10^{47}) \pm (7.15 \times 10^{45})$ (6.9\%) \\
      \hline
      HESS J1731-347 & $20.3 \pm 1.45$ (7.1\%) & $1.64 \pm 0.041$ (2.5\%) & $(1.92 \times 10^{47}) \pm (9.32 \times 10^{45})$ (4.9\%) \\
      \hline
      RX J1713.7-3946 & $74.1 \pm 1.86$ (2.5\%) & $1.75 \pm 0.0067$ (0.39\%) & $(2.24 \times 10^{47}) \pm (2.12 \times 10^{45})$ (1.0\%) \\
      \hline
    \end{tabular}
    \caption{\label{tab:rec_params}The mean and standard deviation for reconstructed parameters across 50 realisations for the four SNRs. The precision $\sigma/\mu$ is also given in brackets.}
\end{table}

To evaluate the accuracy of the simulation, we computed the relative precision based on the resulting mean $\mu$ and standard deviation $\sigma$ of each fitted parameter from each realisation using the formula $\sigma/\mu$. Based on the results summarized in Table~\ref{tab:rec_params}, we note that the spectral index ($\alpha_p$) and amplitude are generally more tightly constrained than the cut-off energy ($E_{\rm p,cut}$). The CTAO simulations demonstrate overall good accuracy in parameter recovery; however, the achievable precision depends on source brightness and spectral hardness. In particular, the precision of the reconstructed cut-off energy degrades when the value lies close to the limits of the energy range within which CTAO is sensitive. In such cases, the statistical limitations become more significant, as can be seen in the cases of Cassiopeia A and HAWC J2227+610. From the plots, we can clearly see that CTAO provides the required VHE sensitivity to constrain the proton cut-off, which is essential for testing PeVatron scenarios, provided that the source has sufficient VHE flux within the CTAO energy range.

\subsection{Constraining physical parameters}

While CTAO will probe VHE $\gamma$-rays, it is crucial to incorporate real lower-energy data available from public archives to gain a more complete understanding of the sources and assess how CTAO observations will improve parameter constraints. MWL studies of these sources also help in constraining the maximum proton energies and distinguishing between leptonic and hadronic emission mechanisms. The real data (RD), as described in the paper MWL23, was added to the simulated CTAO data at high energies to make a combined dataset (CD). This CD was then used for SED fitting with Naima, allowing us to evaluate how the inclusion of CTAO data impacts the derived parameters. To avoid instrumental systematics, any real observation overlapping with the CTAO energy range has been excluded from the analysis. 

\begin{figure}[H]
        \centering
	\includegraphics[scale=0.5]{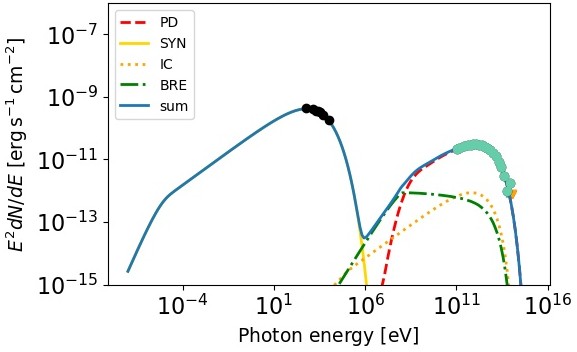}
    \caption{Naima fit result of SNR RX J1713.7-3946 using different theoretical models: pion-decay (PD), synchrotron (SYN), Inverse Compton (IC), bremsstrahlung (BRE). The sum of the different contributions is shown with the blue line.  The real data is shown in black dots, and the CTAO simulated flux is shown in cyan dots. The simulated flux shown corresponds to a single realization of the model.}
    \label{fig:rxj_naima_cta}
\end{figure}

Figure \ref{fig:rxj_naima_cta} shows an example of the fit for the SNR RX J1713.7-3946. The overall MWL fit (solid blue curve) incorporates contributions from synchrotron emission (yellow solid line), bremsstrahlung (green dash–dotted line), inverse Compton scattering (yellow dotted line), and pion-decay (red dashed line).  As can be seen in the figure, the contribution at high energies is mainly dominated by pion-decay emission.

\begin{table}[H]
\centering
\resizebox{1.1\textwidth}{!}{%
\begin{tabular}{ccccccccc}
\hline
Source & $E_{p,\mathrm{cut,\,RD}}$ (TeV) & $E_{p,\mathrm{cut,\,CD}}$ (TeV) & Rel Diff (\%) & $\alpha_{p,\mathrm{RD}}$ & $\alpha_{p,\mathrm{CD}}$ & Rel Diff (\%) \\
\hline
RX J1713.7$-$3946 & $74.10^{+0.04}_{-0.05}$ & $69.5^{+0.02}_{-0.02}$ & $-6.2$ & 1.8 & 1.7 & 5.5  \\
Cassiopeia A & $23.40^{+0.01}_{-0.01}$ & $18.37^{+0.01}_{-0.01}$ & $-21.5$ & 2.1 & 2.0 & -4.76 \\
HESS J1731$-$347 & $20.40^{+0.04}_{-0.03}$ & $18.80^{+0.03}_{-0.03}$ & $-7.8$ & 1.6 & 1.6 & 0.0 \\
HAWC J2227+610 & $446.68^{+0.07}_{-0.07}$ & $323.60^{+0.06}_{-0.08}$ & $-27.6$ & 1.8 & 1.7 & 5.6  \\
\hline
\end{tabular}%
}
\caption{Comparison of reconstructed proton cut-off energies ($E_{p,\mathrm{cut}}$), spectral indices ($\alpha_p$), and likelihood values ($\mathcal{L}$) obtained from the real-data (RD) and combined (CD) datasets. The relative difference indicates the fractional deviation of the CD values from the true injected ones.}
\label{tab:naima_results}
\end{table}

To assess the effect of adding CTAO data, we computed the relative difference between the parameters resulting from the fit of the CD and RD (used as input for simulation from the MWL23 study).
For each SNR, the relative difference (Rel Diff), the proton cut-off energy ($E_{p,\mathrm{cut}}$) values, and proton spectral index ($\alpha_p$), derived from the fit of the CD, along with those derived from real data only, are shown in Table \ref{tab:naima_results}. The results show that the spectral index is consistently recovered to within ~6\%, confirming its robustness. The cut-off energy is generally well constrained up to 100 TeV, with relative deviations below 10\%. However, for sources whose true proton cut-off lies near or beyond the CTAO sensitivity boundaries (either too low or too high), the fitted cut-off tends to be underestimated by up to $\sim$ 30 \%. This underestimation occurs because CTAO detects too few photons above the true cut-off energy, making the spectral turnover difficult to resolve accurately. This indicates that while CTAO will reliably constrain slopes and moderate cut-offs, at the extreme end, it may only provide lower limits on the cut-off energy. HAWC J2227+610 and RX J1713.7−3946 show the largest improvements, likely due to strong VHE emission in the CTAO energy range. Overall, the inclusion of CTAO data at high energies is crucial, as it significantly enhances the constraint on the models, particularly in the multi-TeV domain where current instruments have large uncertainties. 

\section{Study of different cut-off energies}

\begin{figure}[H]
        \centering`
	\includegraphics[width=0.7\columnwidth]{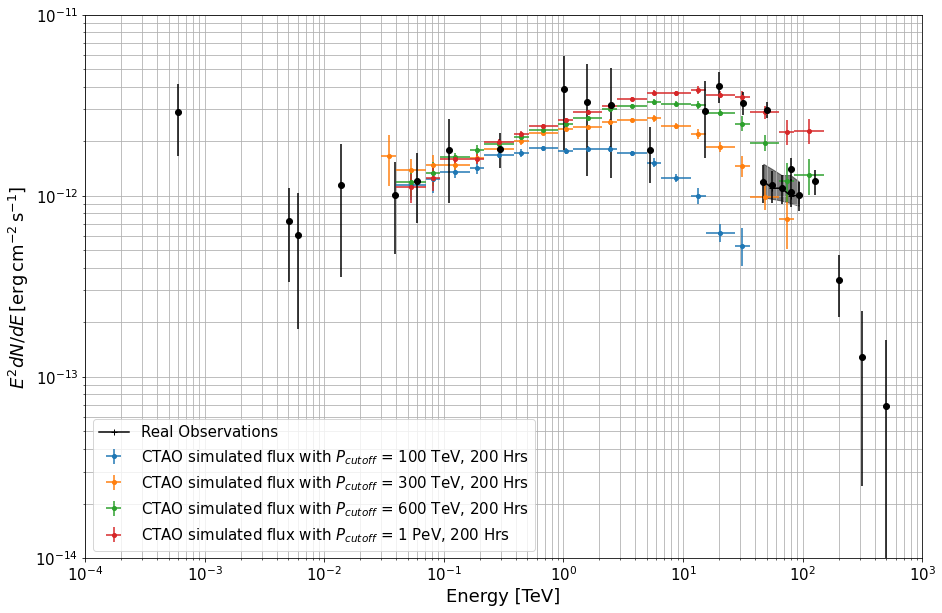}
    \caption{Simulated CTAO fluxes for different proton cut-off energies: $E_{cut,p} = 300, 600,$ and $1000$\,TeV represented by blue, orange, and green data points, respectively. For comparison, the real data for the source are also shown in black and come from \textit{Fermi}-LAT \cite{2019ApJ...885..162X}, VERITAS \cite{Acciari_2009}, LHAASO \cite{2021Natur.594...33C}, and HAWC \cite{Albert_2020}. The simulated flux shown corresponds to a single realization of the model.}
    \label{fig:PD_cut-off}
\end{figure}

The limited sensitivity of current instruments at the highest energies has, until recently, prevented robust constraints on the maximum energy of accelerated protons in Galactic sources. In the case of HAWC J2227+610, both HAWC \citep{Albert_2020} and LHAASO \citep{2021Natur} have detected very-high-energy $\gamma$-ray emission extending up to 10$^{14}$\,eV, suggesting the acceleration of PeV protons. Thus, this source serves as a test case for quantifying CTAO’s sensitivity to proton cut-off energies up to 1 PeV, a key requirement for probing the underlying acceleration mechanisms in the high-energy astrophysical sources.

We have conducted two complementary simulation-based studies 1) evaluating the capability of CTAO to discriminate between different cut-off scenarios for different observation times, and (2) quantifying how the precision of the reconstructed cut-off energy evolves with observation time. The simulations were performed as explained in Section \ref{3.3} by using only the pion-decay model. The ECPL spectral model has been used to define proton distribution with varying cut-off energies: $E_{cut,p}$ = 100, 300, 600, and 1000\,TeV. Three different observation times, 50, 100, and 200 hours, have been considered.

\subsection{Cut-off energy precision study}

To study the precision with which CTAO reconstructs the $E_{cut,p}$ for different cut-off models with varying observation times, for each exposure (50h, 100h, and 200h), 25 realizations were produced. We then fitted each simulated dataset to the model used to generate it. The cut-off energy was left free to vary, while the remaining spectral parameters remained fixed. For each configuration, the mean and standard deviation of the recovered cut-off energies were computed across the 25 realizations, and the relative precision was calculated (defined above). 

\begin{table}[H]
\centering
\resizebox{1.05\textwidth}{!}{%
\begin{tabular}{c|cc|cc|cc}
\hline
\textbf{Injected Model} & \multicolumn{2}{c|}{\textbf{50 h}} & \multicolumn{2}{c|}{\textbf{100 h}} & \multicolumn{2}{c}{\textbf{200 h}} \\
\cline{2-7}
 & Mean $E_\mathrm{cut}$ [TeV] & Precision & Mean $E_\mathrm{cut}$ [TeV] & Precision & Mean $E_\mathrm{cut}$ [TeV] & Precision \\
\hline
PD--100 TeV  & $94.61 \pm 15.16$  & 0.160 & $98.62 \pm 9.05$  & 0.092 & $98.26 \pm 7.24$  & 0.074 \\
PD--300 TeV  & $285.24 \pm 41.58$ & 0.146 & $292.71 \pm 24.23$ & 0.083 & $295.62 \pm 20.87$ & 0.071 \\
PD--600 TeV  & $574.62 \pm 91.06$ & 0.158 & $576.77 \pm 58.61$ & 0.102 & $595.38 \pm 44.80$ & 0.075 \\
PD--1000 TeV & $963.61 \pm 155.43$ & 0.161 & $989.48 \pm 105.22$ & 0.106 & $1004.18 \pm 77.20$ & 0.077 \\
\hline
\end{tabular}%
}
\caption{Reconstructed proton cut-off energies and relative precisions for different simulated livetimes.}
\label{tab:cutoff_precision}
\end{table}

Table \ref{tab:cutoff_precision} contains the resulting mean and precision values for different cut-off models for different observing times. The reconstructed mean cut-off energies show good agreement with the injected value in all cases, with deviations decreasing with increasing observation time. Additionally, we see that the precision systematically increases by a factor of $\sim$ 2 as we move from 50h to 200h for all tested cut-off energies. However, the precision worsens for the highest cut-off values, because for very large 
$E_{p,\mathrm{cut}}$, the pion-decay spectrum remains nearly a power law within the CTAO 
energy range, leaving too few photons in the spectral turnover region to constrain the 
cut-off accurately.

Figure \ref{fig:PD_cut-off} illustrates an example of the simulated CTAO SED for a single realization of a 200 h observation for pion-decay models with different proton cut-off energies of 100, 300, 600, and 1000 TeV, shown by blue, orange, green, and red flux points, respectively. The observed data from Verna et al. \cite{Verna:2021} are shown for comparison, by black datapoints. From the figure, we can see that CTAO flux points have considerably lower uncertainty compared to the observed fluxes in a similar energy range. Additionally, the simulated fluxes start to differ from one another significantly above 1\,TeV. \\

\subsection{Cut-off model discrimination study}

With the flux now simulated, the next step is to assess the capability of CTAO to distinguish between models with different spectral cut-offs. Again, we used the simulated dataset produced by pion-decay models with exponential cut-offs at 100, 300, 600, and 1000 TeV. For each spectral model (hereafter PD–100 TeV, PD–300 TeV, PD–600 TeV, and PD–1000 TeV), we generated N = 25 statistically independent datasets. Each realization was then fitted under the hypothesis of all four candidate cut-off models. During the fitting procedure, the spectral index and normalization of the proton distribution were treated as free parameters, while the cut-off energy was fixed to the value defined by the tested model.

\begin{figure}[H]
        \centering
	\includegraphics[width=0.7\columnwidth]{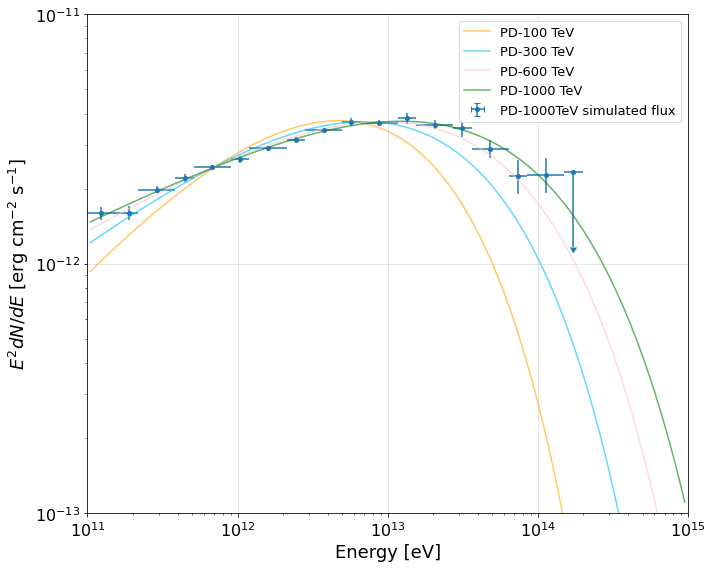}
    \caption{Simulated data for HAWC J2227+610 by using a proton cut-off energy at 300\,TeV (dark blue). The results of the fit with different models (300\,TeV, 600\,TeV and 1\,PeV) are shown in solid lines. The simulated flux shown corresponds to a single realization of the model.}
    \label{fig:PD_300_H0}
\end{figure}
To compare the level of agreement of different hypotheses, say, H0 and H1, with a given dataset, we define the test statistic (TS) as the log-likelihood ratio:

\begin{align}  TS = -2 ln (\frac{L_{H0}}{L_{H1}}). \label{eq:3} \end{align}




For each simulated dataset, we computed the likelihood Test Statistic (TS) differences relative to alternative cut-off hypotheses, and converted these into significances using $\sigma = \sqrt{\Delta TS}$.
As an example, the Figure \ref{fig:PD_300_H0} shows a simulated CTAO SED flux (dark blue) with an injected model with proton cut-off energy of 1000 TeV resulting from 200 h of observations. The simulated flux points are compared to hypotheses of PD–100 TeV, PD–300 TeV, PD–600 TeV, and PD–1000 TeV, shown by orange, blue, pink, and green curves, respectively. The best agreement is found for the 1000 TeV model, indicating that the simulated dataset follows the injected model well. The significant curvature in the alternative models highlights how deviations from the true cut-off lead to visibly poorer fits, demonstrating qualitatively CTAO’s capability to constrain high-energy spectral cut-offs.

\begin{figure}[H]
        \centering
	\includegraphics[width=0.7\columnwidth]{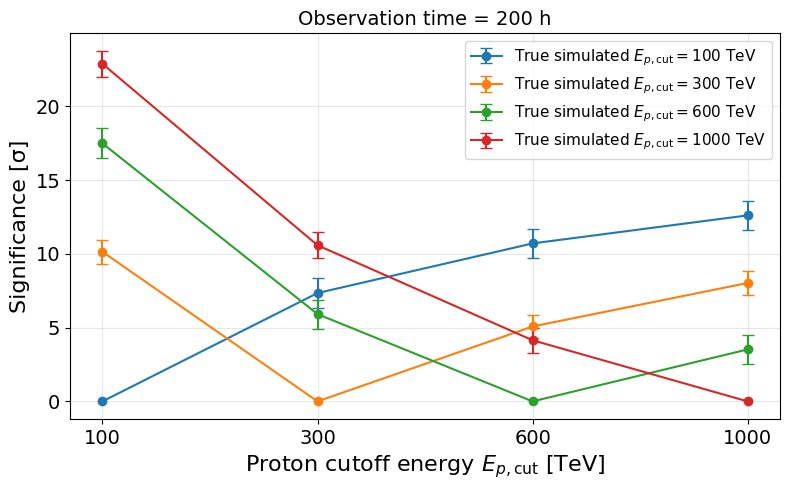}
    \caption{Significance at which the test hypothesis on the proton cut-off energy $E_{p,cut}$ is rejected, for different true injected cut-off values used in the simulations, for an observation time of 200 h. Error bars indicate the variance derived from 25 independent realisations. Each curve corresponds to a different injected (base) model. The PD-100 TeV, PD-300 TeV, PD-600 TeV, and PD-1000 TeV models are represented by blue, orange, green, and red curves, respectively.}
    \label{fig:PD_cutoff_all}
\end{figure}



Figure \ref{fig:PD_cutoff_all} shows the resulting significance obtained when different proton cut-off models are tested against a given “base” dataset with a fixed model. This example is based on 200 h of observation. The PD-100 TeV, PD-300 TeV, PD-600 TeV, and PD-1000 TeV models are represented by blue, orange, green, and red curves, respectively. Each data point shows the statistical preference of the base model compared to alternative cut-off hypotheses. In all cases, the significance drops to its minimum when the tested model corresponds to the same cut-off value used to generate the base dataset. This figure shows that at 200 h, CTAO can reliably distinguish cut-offs separated by a factor of 2 in the 300–600 TeV range, but cannot yet reach 5$\sigma$ for 600 vs 1000 TeV. Therefore, we obtain a discrimination threshold of $\sim$ 600 TeV. However, if smaller observation times are used, this separation becomes more challenging. Moreover, the results of such a cut-off study are strongly dependent on the characteristics of the source. For sources such as HAWC J2227+610, which show promising characteristics for being a PeVatron source, such long dedicated observation times could be chosen.

It is important to note that the ability to distinguish between different proton cut-off energies depends strongly on the shape of the underlying proton spectrum. In scenarios where the particle distribution follows a steep power law with an exponential cut-off, the $\gamma$-ray flux drops rapidly at high energies, even before the exponential suppression dominates. This leads to a suppression of photon statistics in the CTAO’s high-energy regime, causing different cut-off models to produce similar spectral shapes. Consequently, a degeneracy arises between the spectral index and the cut-off energy, limiting the precision with which the latter can be constrained. In such cases, meaningful constraints may require deeper observations, joint analyses with data from other observatories, or informed priors based on MWL observations. This effect should be considered when interpreting the results of our simulations for HAWC J2227+610 and assessing CTAO’s sensitivity to spectral curvature.

\section{Conclusions}
This study explores the potential of CTAO to probe the nature of particle acceleration in selected Galactic SNRs, with a focus on identifying PeVatron candidates. To do so, we have used open software, \textsc{Gammapy (version 0.19)} \cite{2019A&A...625A..10N}, to simulate CTAO flux using radiative models found in the \textsc{Naima} library \cite{Naima_zabalza}. We focused on four potential PeVatron sources, Cassiopeia A, HAWC J2227+610, HESS J1731-347, and RX J1713.7-3946, which showed significant hadronic contributions ranging up to 100\,TeV \cite{SNR_Sharma}. 

First, to ensure that our results reliably reflect the statistical behavior expected in real CTAO observations, we analyzed multiple independent realizations of the simulated spectra for different observation times. The precision study for reconstructed parameters showed that the CTAO simulations demonstrate overall good accuracy in parameter recovery. Additionally, through the relative precision of reconstructed vs injected parameters that the spectral index and amplitude are generally more tightly constrained than the cut-off energy. Furthermore, the simulated CTAO data were compared with existing $\gamma$-ray observations. The flux uncertainties have been reduced by roughly an order of magnitude compared to current instruments. The results showed that the spectral index is consistently recovered to within 6\%, confirming its robustness. However, for sources whose true proton cut-off lies near or beyond the CTAO sensitivity boundaries, the fitted cut-off tends to be underestimated by up to $\sim$ 30\%. 

Second, through MWL SED fitting that includes simulated CTAO observations at high-energy regime, we showed that model parameters can be reliably recovered when they fall within the CTAO energy range. In addition, MWL data are essential for constraining the leptonic emission, ensuring that the inferred hadronic component is not biased. Overall, the inclusion of CTAO data at the highest energies is critical, as it substantially improves model constraints, particularly in the multi-TeV regime where current instruments suffer from large uncertainties.

To examine CTAO’s sensitivity to different proton energy cut-offs, we focused on HAWC J2227+610, as we have observations of the source extending up to the highest energies The main question that we addressed was to estimate the maximum proton cut-off energy at which CTAO can distinguish between different fluxes. To address this question, we first quantified how the precision of reconstructed cut-off energies evolves with observation times. The analysis showed that both accuracy and precision of the cut-off determination improve steadily with increased exposure, with up to a factor of two gain when moving from 50 h to 200 h observations. 

To evaluate CTAO's ability to discriminate between different cut-off models, we performed a cut-off model discrimination study. The model-discrimination test revealed that CTAO can confidently exclude extreme cut-off scenarios (e.g. 100 vs. 1000 TeV) already at 50 h, while 200 h observations enable $\geq 3\sigma$ separation between cut-offs differing by a factor of two in the 300–600 TeV range. Through calculating the TS and significance, we could determine the detection limit for CTAO for proton cut-off energy to be about $\sim 600$\,TeV for the studied source. For sources such as HAWC J2227+610, which show promising characteristics for being a PeVatron source, such long dedicated observation times can be justified.

CTAO’s enhanced angular and energy resolution, along with its improved sensitivity, offer a promising path toward accurately determining the cut-off energies in proton spectra. When complemented by MWL observations from other instruments, this capability can significantly improve our ability to characterize the nature of $\gamma$-ray sources and assess their potential as cosmic-ray accelerators. Nevertheless, the depth and reliability of such studies will strongly depend on the specific source characteristics and will often require long observation times to reach the necessary sensitivity, particularly in the search for PeVatrons.

\section*{Acknowledgements}
\label{Acknowledgements}

We gratefully acknowledge financial support from the agencies and organizations listed here: \url{http://www.cta-observatory.org/consortium_acknowledgments}. 

\noindent
This paper has gone through internal review by the CTAO Consortium.

\noindent
This research has used the instrument response functions provided by the Cherenkov Telescope Array Observatory, \& Cherenkov Telescope Array Consortium. (2021). CTAO Instrument Response Functions - prod5 version v0.1 (v0.1) [Data set]. Zenodo. $https://doi.org/10.5281/zenodo.5499840$

\bibliographystyle{JHEP}
\bibliography{bibliography}

@article{ACERO_pev_CTA,
title = {Sensitivity of the Cherenkov Telescope Array to spectral signatures of hadronic PeVatrons with application to Galactic Supernova Remnants},
journal = {Astroparticle Physics},
volume = {150},
pages = {102850},
year = {2023},
issn = {0927-6505},
doi = {https://doi.org/10.1016/j.astropartphys.2023.102850},
url = {https://www.sciencedirect.com/science/article/pii/S0927650523000361},
author = {F. Acero and et al.}
}

@article{abe2024prospects,
  title={Prospects for a survey of the galactic plane with the Cherenkov Telescope Array},
  author={Abe, S and Abhir, J and Abhishek, A and Acero, F and Acharyya, A and Adam, R and Aguasca-Cabot, A and Agudo, I and Aguirre-Santaella, A and Alfaro, J and others},
  journal={Journal of Cosmology and Astroparticle Physics},
  volume={2024},
  number={10},
  pages={081},
  year={2024},
  publisher={IOP Publishing}
}

@article{CasA_1999,
  title={The bulk expansion of the supernova remnant Cassiopeia A at 151 MHz},
  author={Ag{\"u}eros, Marcel Andre and Green, DA},
  journal={Monthly Notices of the Royal Astronomical Society},
  volume={305},
  number={4},
  pages={957--965},
  year={1999},
  publisher={The Royal Astronomical Society}
}

@article{helder2008characterizing,
  title={Characterizing the nonthermal emission of Cassiopeia A},
  author={Helder, EA and Vink, J},
  journal={The Astrophysical Journal},
  volume={686},
  number={2},
  pages={1094},
  year={2008},
  publisher={IOP Publishing}
}

@article{bao2021hard,
  title={On the Hard Gamma-Ray Spectrum of the Potential PeVatron Supernova Remnant G106. 3+ 2.7},
  author={Bao, Yiwei and Chen, Yang},
  journal={The Astrophysical Journal},
  volume={919},
  number={1},
  pages={32},
  year={2021},
  publisher={IOP Publishing}
}

@article{cui2019snr,
  title={Is the SNR HESS J1731-347 colliding with molecular clouds?},
  author={Cui, Yudong and Yang, Ruizhi and He, Xinbo and Tam, PH Thomas and P{\"u}hlhofer, Gerd},
  journal={The Astrophysical Journal},
  volume={887},
  number={1},
  pages={47},
  year={2019},
  publisher={IOP Publishing}
}

@article{abramowski2011new,
  title={A new SNR with TeV shell-type morphology: HESS J1731-347},
  author={Abramowski, A and Acero, Fabio and Aharonian, F and Akhperjanian, AG and Anton, G and Balzer, Agn{\`e}s and Barnacka, Anna and De Almeida, U Barres and Becherini, Yvonne and Becker, J and others},
  journal={Astronomy \& Astrophysics},
  volume={531},
  pages={A81},
  year={2011},
  publisher={EDP Sciences}
}

@article{tsuji2016expansion,
  title={Expansion measurements of supernova remnant RX J1713. 7- 3946},
  author={Tsuji, Naomi and Uchiyama, Yasunobu},
  journal={Publications of the Astronomical Society of Japan},
  pages={psw102},
  year={2016},
  publisher={Oxford University Press}
}

@article{abdalla2018hess,
  title={HESS observations of RX J1713. 7- 3946 with improved angular and spectral resolution: Evidence for gamma-ray emission extending beyond the X-ray emitting shell},
  author={Abdalla, H and Abramowski, A and Aharonian, F and Benkhali, F Ait and Akhperjanian, AG and Andersson, Tom and Ang{\"u}ner, EO and Arrieta, M and Aubert, P and Backes, M and others},
  journal={Astronomy \& Astrophysics},
  volume={612},
  pages={A6},
  year={2018},
  publisher={EDP Sciences}
}

@article{acharya2017science,
    title = {Science with the Cherenkov telescope array},
    author = {Acharya, Bannanje Sripathi and et al.},
    journal = {arXiv:1709.07997},
    year = {2017},
    publisher = {World Scientific}
}

@misc{SNR_Sharma,
  doi = {10.48550/ARXIV.2207.02695},
  url = {https://arxiv.org/abs/2207.02695},
  author = {Sharma, Pooja and Ou, Ziwei and Henry-Cadrot, Charles and Dubos, Coline and Suomijärvi, Tiina},
  title = {Multiwavelength Analysis of Galactic Supernova Remnants},
  publisher = {arXiv},
  year = {2022},
  copyright = {Creative Commons Attribution Non Commercial Share Alike 4.0 International}
}

@article{Naima_zabalza,
  author = "Zabalza, Victor",
  title = "{Naima: a Python package for inference of particle distribution properties from nonthermal spectra }",
  doi = "10.22323/1.236.0922",
  journal = "PoS",
  year = 2016,
  volume = "ICRC2015",
  pages = "922"
}

@article{Verna:2021,
    author = "Verna, Gaia and Cassol, Franca and Costantini, Heide",
    collaboration = "CTA Consortium",
    title = "{HAWC J2227+610: a potential PeVatron candidate for the CTA in the northern hemisphere}",
    eprint = "2110.07939",
    archivePrefix = "arXiv",
    primaryClass = "astro-ph.HE",
    doi = "10.22323/1.395.0904",
    journal = "PoS",
    volume = "ICRC2021",
    pages = "904",
    year = "2021"
}

@article{James:1975dr,
    author = "James, F. and Roos, M.",
    title = "{Minuit: A System for Function Minimization and Analysis of the Parameter Errors and Correlations}",
    reportNumber = "CERN-DD-75-20",
    doi = "10.1016/0010-4655(75)90039-9",
    journal = "Comput. Phys. Commun.",
    volume = "10",
    pages = "343--367",
    year = "1975"
}

@article{Albert_2020,
   title= "{HAWC J2227+610 and Its Association with G106.3+2.7, a New Potential Galactic PeVatron}",
   volume={896},
   ISSN={2041-8213},
   url={http://dx.doi.org/10.3847/2041-8213/ab96cc},
   DOI={10.3847/2041-8213/ab96cc},
   number={2},
   journal={The Astrophysical Journal},
   publisher={American Astronomical Society},
   author={Albert, A. and Alfaro, R. and Alvarez, C. and Camacho, J. R. Angeles and Arteaga-Velázquez, J. C. and Arunbabu, K. P. and Rojas, D. Avila and Solares, H. A. Ayala and Baghmanyan, V. and Belmont-Moreno, E. and et al.},
   year={2020},
   month={06},
   pages={L29}
}

@ARTICLE{2021Natur,
       author = {{Cao}, Zhen and {Aharonian} et al.},
        title = "{Ultrahigh-energy photons up to 1.4 petaelectronvolts from 12 {\ensuremath{\gamma}}-ray Galactic sources}",
      journal = {Nature},
         year = 2021,
        month = jun,
       volume = {594},
       number = {7861},
        pages = {33-36},
          doi = {10.1038/s41586-021-03498-z},
       adsurl = {https://ui.adsabs.harvard.edu/abs/2021Natur.594...33C}
}

@misc{rxj17132000evidence,
      title={Evidence for TeV gamma-ray emission from the shell type SNR RXJ1713.7-3946}, 
      author={H. Muraishi and et al.},
      year={2000},
      eprint={astro-ph/0001047},
      archivePrefix={arXiv},
      primaryClass={astro-ph}
}

@article{rxj1713_hess_2018,
	author = {{H.E.S.S. Collaboration}},
	title = {H.E.S.S. observations of RX J1713.7−3946 with improved angular and spectral resolution: Evidence for gamma-ray emission extending beyond the X-ray emitting shell},
	DOI= "10.1051/0004-6361/201629790",
	url= "https://doi.org/10.1051/0004-6361/201629790",
	journal = {A\&A},
	year = 2018,
	volume = 612,
	pages = "A6",
}

@article{Acero_2017_rxj1713_cta,
doi = {10.3847/1538-4357/aa6d67},
url = {https://dx.doi.org/10.3847/1538-4357/aa6d67},
year = {2017},
month = {may},
publisher = {The American Astronomical Society},
volume = {840},
number = {2},
pages = {74},
author = {F. Acero and et al.},
title = {Prospects for Cherenkov Telescope Array Observations of the Young Supernova Remnant RX J1713.7−3946},
journal = {The Astrophysical Journal}
}

@article{cassA_gamma_evidence,
	author = {{Aharonian, F.} and et al.}}

@article{Abeysekara_2020_cassA,
doi = {10.3847/1538-4357/ab8310},
url = {https://dx.doi.org/10.3847/1538-4357/ab8310},
year = {2020},
month = {may},
publisher = {The American Astronomical Society},
volume = {894},
number = {1},
pages = {51},
author = {A. U. Abeysekara and et al.},
title = {Evidence for Proton Acceleration up to TeV Energies Based on VERITAS and Fermi-LAT Observations of the Cas A SNR},
journal = {The Astrophysical Journal}
}

@ARTICLE{2011A&A_hess_j1731,
       author = {{H.~E.~S.~S. Collaboration}},
        title = "{A new SNR with TeV shell-type morphology: HESS J1731-347}",
      journal = {A\&A},
         year = 2011,
        month = jul,
       volume = {531},
          eid = {A81},
        pages = {A81},
          doi = {10.1051/0004-6361/201016425},
archivePrefix = {arXiv},
       eprint = {1105.3206},
 primaryClass = {astro-ph.HE},
       adsurl = {https://ui.adsabs.harvard.edu/abs/2011A&A...531A..81H}
}

@ARTICLE{2019_hessj1731_MCs,
       author = {{Cui}, Yudong and et al.},
        title = "{Is the SNR HESS J1731-347 Colliding with Molecular Clouds?}",
      journal = {ApJ},
         year = 2019,
        month = dec,
       volume = {887},
       number = {1},
          eid = {47},
        pages = {47},
          doi = {10.3847/1538-4357/ab4ea0},
archivePrefix = {arXiv},
       eprint = {1904.01761},
 primaryClass = {astro-ph.HE},
       adsurl = {https://ui.adsabs.harvard.edu/abs/2019ApJ...887...47C}
}

@article{articletevcat,
author = {Horan, Deirdre and Wakely, S.},
year = {2008},
month = {03},
pages = {},
title = {TeVCat: An Online Catalog for TeV Astronomy}
}

@ARTICLE{2019A&A...625A..10N,
       author = {{Nigro}, C. and {Deil}, C. and {Zanin}, R. and {Hassan}, T. and {King}, J. and {Ruiz}, J.~E. and {Saha}, L. and {Terrier}, R. and {Br{\"u}gge}, K. and {N{\"o}the}, M. and {Bird}, R. and {Lin}, T.~T.~Y. and {Aleksi{\'c}}, J. and {Boisson}, C. and {Contreras}, J.~L. and {Donath}, A. and {Jouvin}, L. and {Kelley-Hoskins}, N. and {Khelifi}, B. and {Kosack}, K. and {Rico}, J. and {Sinha}, A.},
        title = "{Towards open and reproducible multi-instrument analysis in gamma-ray astronomy}",
      journal = {Astronomy \& Astrophysics},
         year = 2019,
        month = may,
       volume = {625},
          eid = {A10},
        pages = {A10},
          doi = {10.1051/0004-6361/201834938},
archivePrefix = {arXiv},
       eprint = {1903.06621},
 primaryClass = {astro-ph.HE},
       adsurl = {https://ui.adsabs.harvard.edu/abs/2019A&A...625A..10N}
}

@ARTICLE{2014PhRvD..90l3014K,
       author = {{Kafexhiu}, Ervin and {Aharonian}, Felix and {Taylor}, Andrew M. and {Vila}, Gabriela S.},
        title = "{Parametrization of gamma-ray production cross sections for p p interactions in a broad proton energy range from the kinematic threshold to PeV energies}",
      journal = {Physical Review D},
         year = 2014,
        month = dec,
       volume = {90},
       number = {12},
          eid = {123014},
        pages = {123014},
          doi = {10.1103/PhysRevD.90.123014},
archivePrefix = {arXiv},
       eprint = {1406.7369},
 primaryClass = {astro-ph.HE},
       adsurl = {https://ui.adsabs.harvard.edu/abs/2014PhRvD..90l3014K}
}

@article{Abdo_2011,
doi = {10.1088/0004-637X/734/1/28},
url = {https://dx.doi.org/10.1088/0004-637X/734/1/28},
year = {2011},
month = {may},
publisher = {The American Astronomical Society},
volume = {734},
number = {1},
pages = {28},
author = {A. A. Abdo and M. Ackermann and M. Ajello and A. Allafort and L. Baldini and J. Ballet and G. Barbiellini and M. G. Baring and D. Bastieri and R. Bellazzini and B. Berenji and R. D. Blandford and E. D. Bloom and E. Bonamente and A. W. Borgland and A. Bouvier and T. J. Brandt and J. Bregeon and M. Brigida and P. Bruel and R. Buehler and S. Buson and G. A. Caliandro and R. A. Cameron and P. A. Caraveo and J. M. Casandjian and C. Cecchi and S. Chaty and A. Chekhtman and C. C. Cheung and J. Chiang and A. N. Cillis and S. Ciprini and R. Claus and J. Cohen-Tanugi and J. Conrad and S. Corbel and S. Cutini and A. de Angelis and F. de Palma and C. D. Dermer and S. W. Digel and E. do Couto e Silva and P. S. Drell and A. Drlica-Wagner and R. Dubois and D. Dumora and C. Favuzzi and E. C. Ferrara and P. Fortin and M. Frailis and Y. Fukazawa and Y. Fukui and S. Funk and P. Fusco and F. Gargano and D. Gasparrini and N. Gehrels and S. Germani and N. Giglietto and F. Giordano and M. Giroletti and T. Glanzman and G. Godfrey and I. A. Grenier and M.-H. Grondin and S. Guiriec and D. Hadasch and Y. Hanabata and A. K. Harding and M. Hayashida and K. Hayashi and E. Hays and D. Horan and M. S. Jackson and G. Jóhannesson and A. S. Johnson and T. Kamae and H. Katagiri and J. Kataoka and M. Kerr and J. Knödlseder and M. Kuss and J. Lande and L. Latronico and S.-H. Lee and M. Lemoine-Goumard and F. Longo and F. Loparco and M. N. Lovellette and P. Lubrano and G. M. Madejski and A. Makeev and M. N. Mazziotta and J. E. McEnery and P. F. Michelson and R. P. Mignani and W. Mitthumsiri and T. Mizuno and A. A. Moiseev and C. Monte and M. E. Monzani and A. Morselli and I. V. Moskalenko and S. Murgia and M. Naumann-Godo and P. L. Nolan and J. P. Norris and E. Nuss and T. Ohsugi and A. Okumura and E. Orlando and J. F. Ormes and D. Paneque and D. Parent and V. Pelassa and M. Pesce-Rollins and M. Pierbattista and F. Piron and M. Pohl and T. A. Porter and S. Rainò and R. Rando and M. Razzano and O. Reimer and T. Reposeur and S. Ritz and R. W. Romani and M. Roth and H. F.-W. Sadrozinski and P. M. Saz Parkinson and C. Sgrò and D. A. Smith and P. D. Smith and G. Spandre and P. Spinelli and M. S. Strickman and H. Tajima and H. Takahashi and T. Takahashi and T. Tanaka and J. G. Thayer and J. B. Thayer and D. J. Thompson and L. Tibaldo and O. Tibolla and D. F. Torres and G. Tosti and A. Tramacere and E. Troja and Y. Uchiyama and J. Vandenbroucke and V. Vasileiou and G. Vianello and N. Vilchez and V. Vitale and A. P. Waite and P. Wang and B. L. Winer and K. S. Wood and H. Yamamoto and R. Yamazaki and Z. Yang and M. Ziegler},
title = {OBSERVATIONS OF THE YOUNG SUPERNOVA REMNANT RX J1713.7−3946 WITH THE FERMI LARGE AREA TELESCOPE},
journal = {The Astrophysical Journal}
}

@article{2007A&A...464..235A,
   title={Primary particle acceleration above 100 TeV in the shell-type supernova remnant RX J1713.7-3946 with deep HESS observations},
   volume={464},
   ISSN={1432-0746},
   url={http://dx.doi.org/10.1051/0004-6361:20066381},
   DOI={10.1051/0004-6361:20066381},
   number={1},
   journal={Astronomy \& Astrophysics},
   publisher={EDP Sciences},
   author={Aharonian, F. and Akhperjanian, A. G. and Bazer-Bachi, A. R. and Beilicke, M. and Benbow, W. and Berge, D. and Bernlöhr, K. and Boisson, C. and Bolz, O. and Borrel, V. and Braun, I. and Brion, E. and Brown, A. M. and Bühler, R. and Büsching, I. and Carrigan, S. and Chadwick, P. M. and Chounet, L.-M. and Coignet, G. and Cornils, R. and Costamante, L. and Degrange, B. and Dickinson, H. J. and Djannati-Ataï, A. and Drury, L. O’C. and Dubus, G. and Egberts, K. and Emmanoulopoulos, D. and Espigat, P. and Feinstein, F. and Ferrero, E. and Fiasson, A. and Fontaine, G. and Funk, Seb. and Funk, S. and Füßling, M. and Gallant, Y. A. and Giebels, B. and Glicenstein, J. F. and Glück, B. and Goret, P. and Hadjichristidis, C. and Hauser, D. and Hauser, M. and Heinzelmann, G. and Henri, G. and Hermann, G. and Hinton, J. A. and Hoffmann, A. and Hofmann, W. and Holleran, M. and Hoppe, S. and Horns, D. and Jacholkowska, A. and de Jager, O. C. and Kendziorra, E. and Kerschhaggl, M. and Khélifi, B. and Komin, Nu. and Konopelko, A. and Kosack, K. and Lamanna, G. and Latham, I. J. and Le Gallou, R. and Lemière, A. and Lemoine-Goumard, M. and Lohse, T. and Martin, J. M. and Martineau-Huynh, O. and Marcowith, A. and Masterson, C. and Maurin, G. and McComb, T. J. L. and Moulin, E. and de Naurois, M. and Nedbal, D. and Nolan, S. J. and Noutsos, A. and Olive, J.-P. and Orford, K. J. and Osborne, J. L. and Panter, M. and Pelletier, G. and Pita, S. and Pühlhofer, G. and Punch, M. and Ranchon, S. and Raubenheimer, B. C. and Raue, M. and Rayner, S. M. and Reimer, A. and Reimer, O. and Ripken, J. and Rob, L. and Rolland, L. and Rosier-Lees, S. and Rowell, G. and Sahakian, V. and Santangelo, A. and Saugé, L. and Schlenker, S. and Schlickeiser, R. and Schröder, R. and Schwanke, U. and Schwarzburg, S. and Schwemmer, S. and Shalchi, A. and Sol, H. and Spangler, D. and Spanier, F. and Steenkamp, R. and Stegmann, C. and Superina, G. and Tam, P. H. and Tavernet, J.-P. and Terrier, R. and Tluczykont, M. and van Eldik, C. and Vasileiadis, G. and Venter, C. and Vialle, J. P. and Vincent, P. and Völk, H. J. and Wagner, S. J. and Ward, M.},
   year={2006},
   month=nov, pages={235–243} }

@ARTICLE{2019ApJ...885..162X,
       author = {{Xin}, Yuliang and {Zeng}, Houdun and {Liu}, Siming and {Fan}, Yizhong and {Wei}, Daming},
        title = "{VER J2227+608: A Hadronic PeVatron Pulsar Wind Nebula?}",
      journal = {The Astrophysical Journal},
         year = 2019,
        month = nov,
       volume = {885},
       number = {2},
          eid = {162},
        pages = {162},
          doi = {10.3847/1538-4357/ab48ee},
archivePrefix = {arXiv},
       eprint = {1907.04972},
 primaryClass = {astro-ph.HE},
       adsurl = {https://ui.adsabs.harvard.edu/abs/2019ApJ...885..162X}
}

@article{Acciari_2009,
   title={DETECTION OF EXTENDED VHE GAMMA RAY EMISSION FROM G106.3+2.7 WITH VERITAS},
   volume={703},
   ISSN={1538-4357},
   url={hattp://dx.doi.org/10.1088/0004-637X/703/1/L6},
   DOI={10.1088/0004-637x/703/1/l6},
   number={1},
   journal={The Astrophysical Journal},
   publisher={American Astronomical Society},
   author={Acciari, V. A. and Aliu, E. and Arlen, T. and Aune, T. and Bautista, M. and Beilicke, M. and Benbow, W. and Boltuch, D. and Bradbury, S. M. and Buckley, J. H. and Bugaev, V. and Butt, Y. and Byrum, K. and Cannon, A. and Cesarini, A. and Chow, Y. C. and Ciupik, L. and Cogan, P. and Cui, W. and Dickherber, R. and Ergin, T. and Fegan, S. J. and Finley, J. P. and Fortin, P. and Fortson, L. and Furniss, A. and Gall, D. and Gillanders, G. H. and Gotthelf, E. V. and Grube, J. and Guenette, R. and Gyuk, G. and Hanna, D. and Holder, J. and Horan, D. and Hui, C. M. and Humensky, T. B. and Kaaret, P. and Karlsson, N. and Kertzman, M. and Kieda, D. and Konopelko, A. and Krawczynski, H. and Krennrich, F. and Lang, M. J. and LeBohec, S. and Maier, G. and McCann, A. and McCutcheon, M. and Millis, J. and Moriarty, P. and Mukherjee, R. and Ong, R. A. and Otte, A. N. and Pandel, D. and Perkins, J. S. and Pohl, M. and Quinn, J. and Ragan, K. and Reyes, L. C. and Reynolds, P. T. and Roache, E. and Rose, H. J. and Schroedter, M. and Sembroski, G. H. and Smith, A. W. and Steele, D. and Swordy, S. P. and Theiling, M. and Toner, J. A. and Vassiliev, V. V. and Vincent, S. and Wagner, R. G. and Wakely, S. P. and Ward, J. E. and Weekes, T. C. and Weinstein, A. and Weisgarber, T. and Williams, D. A. and Wissel, S. and Wood, M. and Zitzer, B.},
   year={2009},
   month=aug, pages={L6–L9} }

@ARTICLE{2021Natur.594...33C,
       author = {{Cao}, Zhen and {Aharonian}, F.~A. and {An}, Q. and {Axikegu}, Bai, L.~X. and {Bai}, Y.~X. and {Bao}, Y.~W. and {Bastieri}, D. and {Bi}, X.~J. and {Bi}, Y.~J. and {Cai}, H. and {Cai}, J.~T. and {Cao}, Zhe and {Chang}, J. and {Chang}, J.~F. and {Chang}, X.~C. and {Chen}, B.~M. and {Chen}, J. and {Chen}, L. and {Chen}, Liang and {Chen}, Long and {Chen}, M.~J. and {Chen}, M.~L. and {Chen}, Q.~H. and {Chen}, S.~H. and {Chen}, S.~Z. and {Chen}, T.~L. and {Chen}, X.~L. and {Chen}, Y. and {Cheng}, N. and {Cheng}, Y.~D. and {Cui}, S.~W. and {Cui}, X.~H. and {Cui}, Y.~D. and {Dai}, B.~Z. and {Dai}, H.~L. and {Dai}, Z.~G. and {Danzengluobu} and {della Volpe}, D. and {D'Ettorre Piazzoli}, B. and {Dong}, X.~J. and {Fan}, J.~H. and {Fan}, Y.~Z. and {Fan}, Z.~X. and {Fang}, J. and {Fang}, K. and {Feng}, C.~F. and {Feng}, L. and {Feng}, S.~H. and {Feng}, Y.~L. and {Gao}, B. and {Gao}, C.~D. and {Gao}, Q. and {Gao}, W. and {Ge}, M.~M. and {Geng}, L.~S. and {Gong}, G.~H. and {Gou}, Q.~B. and {Gu}, M.~H. and {Guo}, J.~G. and {Guo}, X.~L. and {Guo}, Y.~Q. and {Guo}, Y.~Y. and {Han}, Y.~A. and {He}, H.~H. and {He}, H.~N. and {He}, J.~C. and {He}, S.~L. and {He}, X.~B. and {He}, Y. and {Heller}, M. and {Hor}, Y.~K. and {Hou}, C. and {Hou}, X. and {Hu}, H.~B. and {Hu}, S. and {Hu}, S.~C. and {Hu}, X.~J. and {Huang}, D.~H. and {Huang}, Q.~L. and {Huang}, W.~H. and {Huang}, X.~T. and {Huang}, Z.~C. and {Ji}, F. and {Ji}, X.~L. and {Jia}, H.~Y. and {Jiang}, K. and {Jiang}, Z.~J. and {Jin}, C. and {Kuleshov}, D. and {Levochkin}, K. and {Li}, B.~B. and {Li}, Cong and {Li}, Cheng and {Li}, F. and {Li}, H.~B. and {Li}, H.~C. and {Li}, H.~Y. and {Li}, J. and {Li}, K. and {Li}, W.~L. and {Li}, X. and {Li}, Xin and {Li}, X.~R. and {Li}, Y. and {Li}, Y.~Z. and {Li}, Zhe and {Li}, Zhuo and {Liang}, E.~W. and {Liang}, Y.~F. and {Lin}, S.~J. and {Liu}, B. and {Liu}, C. and {Liu}, D. and {Liu}, H. and {Liu}, H.~D. and {Liu}, J. and {Liu}, J.~L. and {Liu}, J.~S. and {Liu}, J.~Y. and {Liu}, M.~Y. and {Liu}, R.~Y. and {Liu}, S.~M. and {Liu}, W. and {Liu}, Y.~N. and {Liu}, Z.~X. and {Long}, W.~J. and {Lu}, R. and {Lv}, H.~K. and {Ma}, B.~Q. and {Ma}, L.~L. and {Ma}, X.~H. and {Mao}, J.~R. and {Masood}, A. and {Mitthumsiri}, W. and {Montaruli}, T. and {Nan}, Y.~C. and {Pang}, B.~Y. and {Pattarakijwanich}, P. and {Pei}, Z.~Y. and {Qi}, M.~Y. and {Ruffolo}, D. and {Rulev}, V. and {S{\'a}iz}, A. and {Shao}, L. and {Shchegolev}, O. and {Sheng}, X.~D. and {Shi}, J.~R. and {Song}, H.~C. and {Stenkin}, Yu. V. and {Stepanov}, V. and {Sun}, Q.~N. and {Sun}, X.~N. and {Sun}, Z.~B. and {Tam}, P.~H.~T. and {Tang}, Z.~B. and {Tian}, W.~W. and {Wang}, B.~D. and {Wang}, C. and {Wang}, H. and {Wang}, H.~G. and {Wang}, J.~C. and {Wang}, J.~S. and {Wang}, L.~P. and {Wang}, L.~Y. and {Wang}, R.~N. and {Wang}, W. and {Wang}, W. and {Wang}, X.~G. and {Wang}, X.~J. and {Wang}, X.~Y. and {Wang}, Y.~D. and {Wang}, Y.~J. and {Wang}, Y.~P. and {Wang}, Zheng and {Wang}, Zhen and {Wang}, Z.~H. and {Wang}, Z.~X. and {Wei}, D.~M. and {Wei}, J.~J. and {Wei}, Y.~J. and {Wen}, T. and {Wu}, C.~Y. and {Wu}, H.~R. and {Wu}, S. and {Wu}, W.~X. and {Wu}, X.~F. and {Xi}, S.~Q. and {Xia}, J. and {Xia}, J.~J. and {Xiang}, G.~M. and {Xiao}, G. and {Xiao}, H.~B. and {Xin}, G.~G. and {Xin}, Y.~L. and {Xing}, Y. and {Xu}, D.~L. and {Xu}, R.~X. and {Xue}, L. and {Yan}, D.~H. and {Yang}, C.~W. and {Yang}, F.~F. and {Yang}, J.~Y. and {Yang}, L.~L. and {Yang}, M.~J. and {Yang}, R.~Z. and {Yang}, S.~B. and {Yao}, Y.~H. and {Yao}, Z.~G. and {Ye}, Y.~M. and {Yin}, L.~Q. and {Yin}, N. and {You}, X.~H. and {You}, Z.~Y. and {Yu}, Y.~H. and {Yuan}, Q. and {Zeng}, H.~D. and {Zeng}, T.~X. and {Zeng}, W. and {Zeng}, Z.~K. and {Zha}, M. and {Zhai}, X.~X. and {Zhang}, B.~B. and {Zhang}, H.~M. and {Zhang}, H.~Y. and {Zhang}, J.~L. and {Zhang}, J.~W. and {Zhang}, L. and {Zhang}, Li and {Zhang}, L.~X. and {Zhang}, P.~F. and {Zhang}, P.~P. and {Zhang}, R. and {Zhang}, S.~R. and {Zhang}, S.~S. and {Zhang}, X. and {Zhang}, X.~P. and {Zhang}, Yong and {Zhang}, Yi and {Zhang}, Y.~F. and {Zhang}, Y.~L. and {Zhao}, B. and {Zhao}, J. and {Zhao}, L. and {Zhao}, L.~Z. and {Zhao}, S.~P. and {Zheng}, F. and {Zheng}, Y. and {Zhou}, B. and {Zhou}, H. and {Zhou}, J.~N. and {Zhou}, P. and {Zhou}, R. and {Zhou}, X.~X. and {Zhu}, C.~G. and {Zhu}, F.~R. and {Zhu}, H. and {Zhu}, K.~J. and {Zuo}, X.},
        title = "{Ultrahigh-energy photons up to 1.4 petaelectronvolts from 12 {\ensuremath{\gamma}}-ray Galactic sources}",
      journal = {Nature},
         year = 2021,
        month = jun,
       volume = {594},
       number = {7861},
        pages = {33-36},
          doi = {10.1038/s41586-021-03498-z},
       adsurl = {https://ui.adsabs.harvard.edu/abs/2021Natur.594...33C}
}

@article{Yuan_2013,
doi = {10.1088/0004-637X/779/2/117},
year = {2013},
month = {dec},
publisher = {The American Astronomical Society},
volume = {779},
number = {2},
pages = {117},
author = {Yuan, Yajie and Funk, Stefan and Jóhannesson, Gülauger and Lande, Joshua and Tibaldo, Luigi and Uchiyama, Yasunobu},
title = {FERMI LARGE AREA TELESCOPE DETECTION OF A BREAK IN THE GAMMA-RAY SPECTRUM OF THE SUPERNOVA REMNANT CASSIOPEIA A},
journal = {The Astrophysical Journal}
}

@article{Albert_2007,
   title={Observation of VHE γ-rays from Cassiopeia A   with the MAGIC telescope},
   volume={474},
   ISSN={1432-0746},
   url={http://dx.doi.org/10.1051/0004-6361:20078168},
   DOI={10.1051/0004-6361:20078168},
   number={3},
   journal={Astronomy \& Astrophysics},
   publisher={EDP Sciences},
   author={Albert, J. and Aliu, E. and Anderhub, H. and Antoranz, P. and Armada, A. and Baixeras, C. and Barrio, J. A. and Bartko, H. and Bastieri, D. and Becker, J. K. and Bednarek, W. and Berger, K. and Bigongiari, C. and Biland, A. and Bock, R. K. and Bordas, P. and Bosch-Ramon, V. and Bretz, T. and Britvitch, I. and Camara, M. and Carmona, E. and Chilingarian, A. and Coarasa, J. A. and Commichau, S. and Contreras, J. L. and Cortina, J. and Costado, M. T. and Curtef, V. and Danielyan, V. and Dazzi, F. and De Angelis, A. and Delgado, C. and de los Reyes, R. and De Lotto, B. and Domingo-Santamaría, E. and Dorner, D. and Doro, M. and Errando, M. and Fagiolini, M. and Ferenc, D. and Fernández, E. and Firpo, R. and Flix, J. and Fonseca, M. V. and Font, L. and Fuchs, M. and Galante, N. and García-López, R. and Garczarczyk, M. and Gaug, M. and Giller, M. and Goebel, F. and Hakobyan, D. and Hayashida, M. and Hengstebeck, T. and Herrero, A. and Höhne, D. and Hose, J. and Hsu, C. C. and Jacon, P. and Jogler, T. and Kosyra, R. and Kranich, D. and Kritzer, R. and Laille, A. and Lindfors, E. and Lombardi, S. and Longo, F. and López, J. and López, M. and Lorenz, E. and Majumdar, P. and Maneva, G. and Mannheim, K. and Mansutti, O. and Mariotti, M. and Martínez, M. and Mazin, D. and Merck, C. and Meucci, M. and Meyer, M. and Miranda, J. M. and Mirzoyan, R. and Mizobuchi, S. and Moralejo, A. and Nilsson, K. and Ninkovic, J. and Oña-Wilhelmi, E. and Otte, N. and Oya, I. and Paneque, D. and Panniello, M. and Paoletti, R. and Paredes, J. M. and Pasanen, M. and Pascoli, D. and Pauss, F. and Pegna, R. and Persic, M. and Peruzzo, L. and Piccioli, A. and Poller, M. and Puchades, N. and Prandini, E. and Raymers, A. and Rhode, W. and Ribó, M. and Rico, J. and Rissi, M. and Robert, A. and Rügamer, S. and Saggion, A. and Sánchez, A. and Sartori, P. and Scalzotto, V. and Scapin, V. and Schmitt, R. and Schweizer, T. and Shayduk, M. and Shinozaki, K. and Shore, S. N. and Sidro, N. and Sillanpää, A. and Sobczynska, D. and Stamerra, A. and Stark, L. S. and Takalo, L. and Temnikov, P. and Tescaro, D. and Teshima, M. and Tonello, N. and Torres, D. F. and Turini, N. and Vankov, H. and Vitale, V. and Wagner, R. M. and Wibig, T. and Wittek, W. and Zandanel, F. and Zanin, R. and Zapatero, J.},
   year={2007},
   month=oct, pages={937–940} }

@article{Acciari_2010,
doi = {10.1088/0004-637X/714/1/163},
url = {https://dx.doi.org/10.1088/0004-637X/714/1/163},
year = {2010},
month = {apr},
publisher = {The American Astronomical Society},
volume = {714},
number = {1},
pages = {163},
author = {Acciari, V. A. and Aliu, E. and Arlen, T. and Aune, T. and Bautista, M. and Beilicke, M. and Benbow, W. and Boltuch, D. and Bradbury, S. M. and Buckley, J. H. and Bugaev, V. and Butt, Y. and Byrum, K. and Cannon, A. and Cesarini, A. and Chow, Y. C. and Ciupik, L. and Cogan, P. and Cui, W. and Dickherber, R. and Duke, C. and Ergin, T. and Fegan, S. J. and Finley, J. P. and Finnegan, G. and Fortin, P. and Fortson, L. and Furniss, A. and Galante, N. and Gall, D. and Gillanders, G. H. and Grube, J. and Guenette, R. and Gyuk, G. and Hanna, D. and Holder, J. and Huang, D. and Hui, C. M. and Humensky, T. B. and Kaaret, P. and Karlsson, N. and Kertzman, M. and Kieda, D. and Konopelko, A. and Krawczynski, H. and Krennrich, F. and Lang, M. J. and LeBohec, S. and Maier, G. and McArthur, S. and McCann, A. and McCutcheon, M. and Millis, J. and Moriarty, P. and Ong, R. A. and Pandel, D. and Perkins, J. S. and Pohl, M. and Quinn, J. and Ragan, K. and Reynolds, P. T. and Roache, E. and Rose, H. J. and Schroedter, M. and Sembroski, G. H. and Smith, A. W. and Smith, B. R. and Steele, D. and Swordy, S. P. and Theiling, M. and Thibadeau, S. and Varlotta, A. and Vassiliev, V. V. and Vincent, S. and Wagner, R. G. and Wakely, S. P. and Ward, J. E. and Weekes, T. C. and Weinstein, A. and Weisgarber, T. and Wissel, S. and Wood, M.},
title = {OBSERVATIONS OF THE SHELL-TYPE SUPERNOVA REMNANT CASSIOPEIA A AT TeV ENERGIES WITH VERITAS},
journal = {The Astrophysical Journal}
}

@article{2011,
   title={A new SNR with TeV shell-type morphology: HESS J1731-347},
   volume={531},
   ISSN={1432-0746},
   url={http://dx.doi.org/10.1051/0004-6361/201016425},
   DOI={10.1051/0004-6361/201016425},
   journal={Astronomy \&; Astrophysics},
   publisher={EDP Sciences},
   author={Abramowski, A. and Acero, F. and Aharonian, F. and Akhperjanian, A. G. and Anton, G. and Balzer, A. and Barnacka, A. and Barres de Almeida, U. and Becherini, Y. and Becker, J. and Behera, B. and Bernlöhr, K. and Bochow, A. and Boisson, C. and Bolmont, J. and Bordas, P. and Brucker, J. and Brun, F. and Brun, P. and Bulik, T. and Büsching, I. and Carrigan, S. and Casanova, S. and Cerruti, M. and Chadwick, P. M. and Charbonnier, A. and Chaves, R. C. G. and Cheesebrough, A. and Chounet, L.-M. and Clapson, A. C. and Coignet, G. and Cologna, G. and Conrad, J. and Dalton, M. and Daniel, M. K. and Davids, I. D. and Degrange, B. and Deil, C. and Dickinson, H. J. and Djannati-Ataï, A. and Domainko, W. and Drury, L.O’C. and Dubois, F. and Dubus, G. and Dutson, K. and Dyks, J. and Dyrda, M. and Egberts, K. and Eger, P. and Espigat, P. and Fallon, L. and Farnier, C. and Fegan, S. and Feinstein, F. and Fernandes, M. V. and Fiasson, A. and Fontaine, G. and Förster, A. and Füßling, M. and Gallant, Y. A. and Gast, H. and Gérard, L. and Gerbig, D. and Giebels, B. and Glicenstein, J. F. and Glück, B. and Goret, P. and Göring, D. and Häffner, S. and Hague, J. D. and Hampf, D. and Hauser, M. and Heinz, S. and Heinzelmann, G. and Henri, G. and Hermann, G. and Hinton, J. A. and Hoffmann, A. and Hofmann, W. and Hofverberg, P. and Holler, M. and Horns, D. and Jacholkowska, A. and de Jager, O. C. and Jahn, C. and Jamrozy, M. and Jung, I. and Kastendieck, M. A. and Katarzyński, K. and Katz, U. and Kaufmann, S. and Keogh, D. and Khangulyan, D. and Khélifi, B. and Klochkov, D. and Kluźniak, W. and Kneiske, T. and Komin, Nu. and Kosack, K. and Kossakowski, R. and Laffon, H. and Lamanna, G. and Lennarz, D. and Lohse, T. and Lopatin, A. and Lu, C.-C. and Marandon, V. and Marcowith, A. and Masbou, J. and Maurin, D. and Maxted, N. and McComb, T. J. L. and Medina, M. C. and Méhault, J. and Moderski, R. and Moulin, E. and Naumann, C. L. and Naumann-Godo, M. and de Naurois, M. and Nedbal, D. and Nekrassov, D. and Nguyen, N. and Nicholas, B. and Niemiec, J. and Nolan, S. J. and Ohm, S. and de Oña Wilhelmi, E. and Opitz, B. and Ostrowski, M. and Oya, I. and Panter, M. and Paz Arribas, M. and Pedaletti, G. and Pelletier, G. and Petrucci, P.-O. and Pita, S. and Pühlhofer, G. and Punch, M. and Quirrenbach, A. and Raue, M. and Rayner, S. M. and Reimer, A. and Reimer, O. and Renaud, M. and de los Reyes, R. and Rieger, F. and Ripken, J. and Rob, L. and Rosier-Lees, S. and Rowell, G. and Rudak, B. and Rulten, C. B. and Ruppel, J. and Ryde, F. and Sahakian, V. and Santangelo, A. and Schlickeiser, R. and Schöck, F. M. and Schulz, A. and Schwanke, U. and Schwarzburg, S. and Schwemmer, S. and Sikora, M. and Skilton, J. L. and Sol, H. and Spengler, G. and Stawarz, Ł. and Steenkamp, R. and Stegmann, C. and Stinzing, F. and Stycz, K. and Sushch, I. and Szostek, A. and Tavernet, J.-P. and Terrier, R. and Tluczykont, M. and Valerius, K. and van Eldik, C. and Vasileiadis, G. and Venter, C. and Vialle, J. P. and Viana, A. and Vincent, P. and Völk, H. J. and Volpe, F. and Vorobiov, S. and Vorster, M. and Wagner, S. J. and Ward, M. and White, R. and Wierzcholska, A. and Zacharias, M. and Zajczyk, A. and Zdziarski, A. A. and Zech, A. and Zechlin, H.-S.},
   year={2011},
   month=jun, pages={A81} }

@ARTICLE{2013_Blasi,
       author = {{Blasi}, Pasquale},
        title = "{The origin of galactic cosmic rays}",
      journal = {Nuclear Physics B Proceedings Supplements},
         year = 2013,
        month = nov,
       volume = {21},
          eid = {70},
        pages = {70},
          doi = {10.1007/s00159-013-0070-7},
archivePrefix = {arXiv},
       eprint = {1311.7346},
 primaryClass = {astro-ph.HE},
       adsurl = {https://ui.adsabs.harvard.edu/abs/2013A&ARv..21...70B}
}

@article{hess2016acceleration,
  title={Acceleration of petaelectronvolt protons in the Galactic Centre},
  journal={Nature},
  volume={531},
  number={7595},
  pages={476--479},
  year={2016},
  publisher={Nature Publishing Group UK London}
}

@article{malkov2001nonlinear,
  title={Nonlinear theoryof diffusive acceleration of particles by shock waves},
  author={Malkov, MA and Drury, L O'C},
  journal={Reports on Progress in Physics},
  volume={64},
  number={4},
  pages={429},
  year={2001},
  publisher={IOP Publishing}
}

@article{kelner2006energy,
  title={Energy spectra of gamma rays, electrons, and neutrinos produced at proton-proton interactions in the very high energy regime},
  author={Kelner, Stanislav R and Aharonian, Felex A and Bugayov, Vistcheslav V},
  journal={Physical Review D—Particles, Fields, Gravitation, and Cosmology},
  volume={74},
  number={3},
  pages={034018},
  year={2006},
  publisher={APS}
}

@article{cristofari2021hunt,
  title={The hunt for pevatrons: The case of supernova remnants},
  author={Cristofari, Pierre},
  journal={Universe},
  volume={7},
  number={9},
  pages={324},
  year={2021},
  publisher={MDPI}
}

@inproceedings{lopez2025ctao,
  title={CTAO status and perspective},
  author={L{\'o}pez-Oramas, Alicia},
  booktitle={EPJ Web of Conferences},
  volume={319},
  pages={01002},
  year={2025},
  organization={EDP Sciences}
}

@article{abe2023multiwavelength,
  title={Multiwavelength study of the galactic PeVatron candidate LHAASO J2108+ 5157},
  author={Abe, S and Aguasca-Cabot, A and Agudo, I and Crespo, N Alvarez and Antonelli, LA and Aramo, C and Arbet-Engels, A and Artero, M and Asano, K and Aubert, P and others},
  journal={Astronomy \& astrophysics},
  volume={673},
  pages={A75},
  year={2023},
  publisher={EDP Sciences}
}

@article{kilpatrick2014interaction,
  title={Interaction between Cassiopeia A and nearby molecular clouds},
  author={Kilpatrick, Charles D and Bieging, John H and Rieke, George H},
  journal={The Astrophysical Journal},
  volume={796},
  number={2},
  pages={144},
  year={2014},
  publisher={IOP Publishing}
}

@article{Foreman_Mackey_2013,
   title={<tt>emcee</tt>: The MCMC Hammer},
   volume={125},
   ISSN={1538-3873},
   url={http://dx.doi.org/10.1086/670067},
   DOI={10.1086/670067},
   number={925},
   journal={Publications of the Astronomical Society of the Pacific},
   publisher={IOP Publishing},
   author={Foreman-Mackey, Daniel and Hogg, David W. and Lang, Dustin and Goodman, Jonathan},
   year={2013},
   month=mar, pages={306–312} }

@article{2021,
   title={Potential PeVatron supernova remnant G106.3+2.7 seen in the highest-energy gamma rays},
   volume={5},
   ISSN={2397-3366},
   url={http://dx.doi.org/10.1038/s41550-020-01294-9},
   DOI={10.1038/s41550-020-01294-9},
   number={5},
   journal={Nature Astronomy},
   publisher={Springer Science and Business Media LLC},
   author={Amenomori, M. and Bao, Y. W. and Bi, X. J. and Chen, D. and Chen, T. L. and Chen, W. Y. and Chen, Xu and Chen, Y. and Cirennima and Cui, S. W. and Danzengluobu and Ding, L. K. and Fang, J. H. and Fang, K. and Feng, C. F. and Feng, Zhaoyang and Feng, Z. Y. and Gao, Qi and Gou, Q. B. and Guo, Y. Q. and Guo, Y. Y. and He, H. H. and He, Z. T. and Hibino, K. and Hotta, N. and Hu, Haibing and Hu, H. B. and Huang, J. and Jia, H. Y. and Jiang, L. and Jin, H. B. and Kasahara, K. and Katayose, Y. and Kato, C. and Kato, S. and Kawata, K. and Kihara, W. and Ko, Y. and Kozai, M. and Labaciren and Le, G. M. and Li, A. F. and Li, H. J. and Li, W. J. and Lin, Y. H. and Liu, B. and Liu, C. and Liu, J. S. and Liu, M. Y. and Liu, W. and Lou, Y.-Q. and Lu, H. and Meng, X. R. and Munakata, K. and Nakada, H. and Nakamura, Y. and Nanjo, H. and Nishizawa, M. and Ohnishi, M. and Ohura, T. and Ozawa, S. and Qian, X. L. and Qu, X. B. and Saito, T. and Sakata, M. and Sako, T. K. and Shao, J. and Shibata, M. and Shiomi, A. and Sugimoto, H. and Takano, W. and Takita, M. and Tan, Y. H. and Tateyama, N. and Torii, S. and Tsuchiya, H. and Udo, S. and Wang, H. and Wu, H. R. and Xue, L. and Yamamoto, Y. and Yang, Z. and Yokoe, Y. and Yuan, A. F. and Zhai, L. M. and Zhang, H. M. and Zhang, J. L. and Zhang, X. and Zhang, X. Y. and Zhang, Y. and Zhang, Yi and Zhang, Ying and Zhao, S. P. and Zhaxisangzhu and Zhou, X. X.},
   year={2021},
   month=mar, pages={460–464} }

@article{condon2017detection,
  title={Detection of two TEV shell-type remnants at GEV energies with FERMI LAT: HESS j1731-347 and SN 1006},
  journal={The Astrophysical Journal},
  volume={851},
  number={2},
  pages={100},
  year={2017},
  publisher={IOP Publishing}
}

@dataset{ctao_IRF,
  author       = {Cherenkov Telescope Array Observatory and
                  Cherenkov Telescope Array Consortium},
  title        = {CTAO Instrument Response Functions - prod5 version v0.1},
  month        = sep,
  year         = 2021,
  publisher    = {Zenodo},
  version      = {v0.1},
  doi          = {10.5281/zenodo.5499840},
  url          = {https://doi.org/10.5281/zenodo.5499840},
}

\end{document}